\documentclass{aa}
\bibpunct{(}{)}{;}{a}{}{,} 
\usepackage{graphicx}
\usepackage{subfig}
\usepackage{hyperref}
\usepackage{comment}
\usepackage{multirow}
\usepackage{xcolor}
\usepackage[normalem]{ulem}

\title{Spinning-down RU~Lup}
\subtitle{Constraints on the physics of the outflow from high-resolution spectroscopy\thanks{Based on observations collected at the European Southern Observatory under ESO programmes 69.C-0481(A), 106.20Z8.003, and 106.20Z8.007.}}
\author{A. Armeni\inst{\ref{inst1}, \ref{inst2}} \and B. Stelzer\inst{\ref{inst1}} \and A. Frasca\inst{\ref{inst3}} \and C.~F. Manara\inst{\ref{inst4}} \and 
J. Campbell-White\inst{\ref{inst4}} \and J.~F.~Gameiro\inst{\ref{inst5},\ref{inst6}} \and M.~Gangi\inst{\ref{inst7},\ref{inst8}} }

\institute{
    Institut für Astronomie und Astrophysik, Eberhard Karls Universität Tübingen, Sand 1, 72076 Tübingen, Germany
    \label{inst1}
    \and 
    INAF – Osservatorio Astronomico di Capodimonte, via Moiariello 16, 80131 Napoli, Italy \\
    \email{antonio.armeni@inaf.it} 
    \label{inst2}
    \and
    INAF – Osservatorio Astrofisico di Catania, via S. Sofia 78, 95123 Catania, Italy
    \label{inst3}
    \and
    European Southern Observatory, Karl-Schwarzschild-Strasse 2, 85748 Garching bei München, Germany
    \label{inst4}
    \and
    Instituto de Astrofísica e Ciências do Espaço, Universidade do Porto, CAUP, Rua das Estrelas, P-4150-762 Porto, Portugal
    \label{inst5}
    \and
    Departamento de Física e Astronomia, Faculdade de Ciências, Universidade do Porto, Rua do Campo Alegre 687, P-4169-007 Porto, Portugal
    \label{inst6}
    \and
    INAF – Osservatorio Astronomico di Roma, via Frascati 33, 00078 Monte Porzio Catone, Italy
    \label{inst7}
    \and
    ASI, Italian Space Agency, via del Politecnico snc, 00133 Roma, Italy
    \label{inst8}
}

\date{26/10/2025}

\abstract
{Magnetic winds are a key mechanism for angular momentum removal in young stars.}{We aim at characterizing the multi-component outflow of RU~Lup, linking discrete absorption components in resonance lines to forbidden-line emission, and quantify the mass loading, lever arms, and torque carried by the wind.}{The unprecedented high resolution of the \textit{Echelle SPectrograph for Rocky Exoplanets and Stable Spectroscopic Observations} enabled a detailed study of the forbidden emission lines and the blueshifted absorption in the lines of the \ion{Na}{i} and \ion{Ca}{ii} doublets, which we resolved in three discrete absorption components {at low, medium, and high velocities. We developed a method that disentangles vertical and toroidal velocities in the absorption components and infers the launching radius $r_0$, magnetic lever arm $\lambda$, and $\dot{M}_{\rm wind}$.}}{We identified a low-velocity broad component in the [\ion{O}{i}] 5577 line, consistent with a rotating magnetohydrodynamic disk wind launched near the disk truncation radius. We showed that the discrete absorption components trace spatially and physically distinct regions of the outflow. The high velocity absorption component is connected to the high velocity component of the forbidden lines and it is formed in a knot in the large-scale, low-density jet. The medium and low velocity components are launched from the inner disk ($r_0 \lesssim 6.76\,R_\star$) with low lever arms indicative of warm, highly mass-loaded streamlines. The two components differ mainly in vertical velocity. The low velocity absorption is consistent with an outer absorbing shell, while the medium velocity absorption forms near the disk truncation radius. Its higher vertical velocity is compatible with either a slightly larger lever arm, or additional heating at the base of the flow. For plausible ionization levels in the inner disk, this outflow component removes a substantial fraction of the accretion spin-up torque.}
{RU~Lup hosts a stratified, rotating, warm disk wind launched across a narrow annulus near the disk truncation radius, which is sufficiently mass-loaded to extract a large amount of the stellar spin-up torque. The observations disfavor an X-wind scenario.}


\begin{document}
    \titlerunning{Spinning-down RU Lup}
    \authorrunning{A. Armeni et al.}
    \maketitle
    \section{Introduction}
    Classical T Tauri Stars (CTTSs) are low-mass ($\lesssim 2 ~ M_{\odot}$) stars in the early ($\sim 1-10$~Myr) stages of their formation. They accrete matter from a circumstellar disk through a magnetosphere \citep{Bouvier+2007, Hartmann+2016}. The evolution of these young stellar objects (YSOs) and the formation of their planetary systems are fundamentally shaped by the interplay between accretion and mass loss. 

    The removal of angular momentum from the accretion disk is essential for disk material to move inward and ultimately accrete onto the star. Turbulence driven by the magnetorotational instability \citep[MRI,][]{BalbusHawley1991} has been found to be largely ineffective in transporting angular momentum outward in protostellar disks, primarily due to the expected low ionization fraction \citep[e.g.,][]{Hartmann2008}. As a result, magnetohydrodynamic (MHD) disk winds have emerged as the leading mechanism for extracting angular momentum from the accretion disk \citep{Lesur2021, Pascucci+2023}.
    
    Another key problem in star formation is how the star counteracts the natural spin-up caused by accretion. The strong ($\sim 1$~kG) magnetic field of CTTSs can truncate the disk at a few stellar radii, and the accreting matter transfers angular momentum to the star \citep[e.g.,][]{MattPudritz2005b, MattPudritz2005a}. The typical angular momentum per unit time transferred from the disk to the star is sufficient to spin it up to break-up velocity in much less than $\sim 1$~Myr \citep{HartmannStauffer1989}. However, CTTSs typically rotate at velocities an order of magnitude lower \citep{Bouvier+2014}. This discrepancy requires a spin-down mechanism, that is, a process that carries away excess angular momentum and regulates the stellar rotation rate. 
    
    Observationally, forbidden emission lines are well-known tracers of the outflowing gas in YSOs {\citep[e.g.,][]{Edwards+1987, Ray+2007, ErcolanoPascucci2017, Banzatti+2019, Sperling+2024}}. These lines typically show two distinct velocity components: a high velocity component (HVC) with blueshifted emission that can reach hundreds of $\rm{km~s}^{-1}$, and a low velocity component (LVC) with blueshifted velocities less than $\sim 50 ~ \rm{km~s}^{-1}$ \citep{Hartigan+1995}.
    The HVC is thought to originate in a jet that is collimated by magnetic fields \citep{KwanTademaru1988} and typically extends to hundreds AU \citep{Hirth+1997}. The LVC originates in a more compact region than the HVC and it is resolved into broad (BC) and narrow (NC) kinematic components \citep{Rigliaco+2013, Simon+2016, Fang+2018}. The large widths of the LVC–BCs point to an MHD disk wind origin at launching radii $\lesssim 0.5$~AU \citep[e.g.,][]{Campbell-White+2023}. The LVC–NC is likely formed at larger disk radii either in a photoevaporative thermal wind \citep{Weber+2020} or in an MHD wind from the outer disk.

    This work focuses on RU~Lup (Sz 83), a young K7 star \citep{Alcala+2017} located in the Lupus 2 cloud at a distance of $158.9 \pm 0.7$~pc \citep{GaiaDR3}. 
    In \citet[][hereafter Paper I]{Armeni+2024}, we analyzed the accretion dynamics of RU Lup, constraining its accretion rate and the location of the disk truncation radius.  
    However, to fully understand the angular momentum evolution of RU~Lup, it is crucial to also characterize its outflows.

    {Estimates of the inclination of RU Lup indicate a nearly face-on orientation. Interferometric observations with the GRAVITY instrument suggest an inclination of about $16-20^{\circ}$, based on the geometry of the inner disk \citep{GravityColl+2021}. From spectroscopy, we inferred an inclination of $i_{\star} = 16 \pm 5 ~^{\circ}$ for the stellar rotation axis from a measure of the projected rotational velocity, $v \sin i$ (Paper\,I) and the stellar rotation period obtained by \citet{Stempels+2007}, that is,  $P_{\star} = 3.71$~days.} 

    The profiles of the {forbidden lines} of RU~Lup exhibit a complex structure \citep{Natta+2014}, and high-resolution spectroscopy enables a more detailed examination of these profiles compared to lower-resolution observations \citep[e.g.,][]{Nisini+2018}.
    Recent studies have provided significant insights into the outflow mechanisms of RU~Lup. \citet{Whelan+2021} performed a spectro-astrometric analysis of the {forbidden lines} in the spectrum of RU~Lup with the \textit{Ultraviolet and Visual Echelle Spectrograph} \citep[UVES,][]{Dekker+2000}.
    Their findings revealed that the LVC-NC of the {forbidden lines} traces wide-angled MHD disk wind originating from the circumstellar disk of RU~Lup.
    Building on these results, \citet{Birney+2024} imaged the outflow of RU~Lup with the \textit{Multi Unit Spectroscopic Explorer} \citep[MUSE,][]{Bacon+2010}, providing information on the spatial width and collimation of the components. They showed that the MHD disk wind traced by the LVC-NC has a full opening angle ($\theta$) of $\sim 37^{\rm o}$, while the jet is more collimated, with $\theta \sim 25^{\rm o}$.
    
    The ability of integrated field spectroscopy to spatially resolve the emission regions makes it particularly useful for tracing the large-scale structure of YSO outflows.     
    However, to study the physics of the wind close to the star, we need to resort to spectroscopy. For this reason, in this work we analyze
    the high-resolution spectra presented in Paper\,I with a focus on the physics of the RU~Lup outflow close to the launching region. 
    Our aim is to find a mechanism that is capable of maintaining the star in spin equilibrium.

    This paper is organized as follows. In Sect.~\ref{obs}, we introduce the observations. We present the analysis of the high resolution spectra in Sect.~\ref{FELs}, where we analyze the {forbidden lines}, and Sect.~\ref{DACs}, where we study the absorption from the wind. We discuss the results of our analysis and the implications for the structure of the outflow in Sect.~\ref{discussion}, followed by a summary of our conclusions in Sect.~\ref{conclusions}.

    \section{Observations}
    \label{obs}
    High resolution ($R = 140000$) optical ($3800 - 7880$~{\AA}) spectra were obtained with the \textit{Echelle SPectrograph for Rocky Exoplanets and Stable Spectroscopic Observations} \citep[ESPRESSO,][]{Pepe+2021} in the framework of the PENELLOPE program \citep{Manara+2021}. {PENELLOPE is part of the ODYSSEUS large program, which complements the \textit{Ultraviolet Legacy Library of Young Stars as Essential Standards} \citep[ULLYSES][]{RomanDuval+2020, Espaillat+2022}, a program carried out with the Hubble Space Telescope (HST).}   
    Two spectra were obtained in 2021 in Pr. Id. 106.20Z8.003 and five in 2022 in Pr. Id. 106.20Z8.007 (PI Manara). 
    The ESPRESSO spectra were reduced by the PENELLOPE team as described by \citet{Manara+2021}. Telluric correction was performed using the molecfit tool \citep{Smette+2015}.
    ESPRESSO is a fiber-fed spectrograph, and the aperture on the sky of the fiber was $1^{\prime\prime}$ during the observations. At the distance of RU~Lup, this limits the size of the observing region to $\sim 160$~AU.
    
    As in Paper\,I, throughout this article each spectrum is labelled as “ID yy.j” where ID is an identification for the spectrograph, yy are the last two digits of the year and j, when needed, is the jth observation from that spectrograph in that year. For example, the 3rd ESPRESSO observation from 2022 is called ES 22.3.
    Table~\ref{tab:log_specobs} reports the log of the spectroscopic observations.

    \section{Forbidden emission lines}
    \label{FELs}
    
    \begin{figure*}
        \centering
        \includegraphics[width=\linewidth]{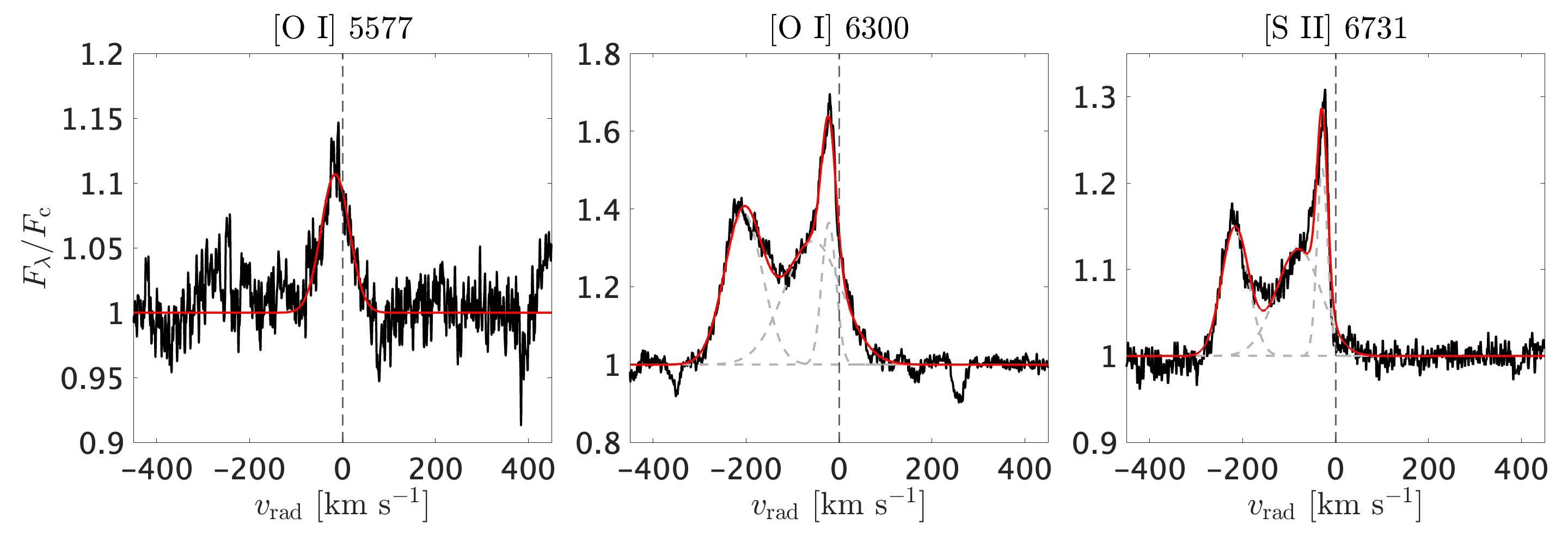}
        \caption{Photospheric subtracted profiles of the [\ion{O}{i}] 5577, [\ion{O}{i}] 6300, and [\ion{S}{ii}] 6731 lines in the ES~22.5 spectrum of RU~Lup, {and best-fitting combinations of Gaussian functions.}}
        \label{fig:FELs_superposed}
    \end{figure*}

    Figure~\ref{fig:FELs_superposed} shows three prominent {forbidden lines} in the ES~22.5 spectrum of RU~Lup, the [\ion{O}{i}] 5577, [\ion{O}{i}] 6300, and [\ion{S}{ii}] 6731 lines. Given their different critical electron densities, $n_{\rm c}$, these lines potentially probe distinct regions of the outflow {\citep[e.g.,][]{OsterbrockFerland2006}}. The values are $n_{\rm c} = 10^8 ~ \rm{cm}^{-3}$, $n_{\rm c} = 1.8 \times 10^6 ~ \rm{cm}^{-3}$, and $n_{\rm c} = 1.6 \times 10^4 ~ \rm{cm}^{-3}$, respectively.
    We removed the photospheric spectrum in the {forbidden lines} as explained in Appendix~\ref{photospheric_subtraction}, {and fitted the line profiles with a combination of Gaussian functions. We used a single Gaussian for the [\ion{O}{i}] 5577 line and three Gaussians for the [\ion{O}{i}] 6300 and [\ion{S}{ii}] 6731 lines.} {The results of the fits are reported in Table~\ref{tab:FELs_gaussian} and shown in Fig.~\ref{fig:FELs_superposed}}.

    \begin{table*}[t]
        \centering
        \caption{{Results of the Gaussian fits to the forbidden emission lines in the ES~22.5 spectrum of RU Lup.}}
        \begin{tabular}{lcccccc}
        \hline
        Line & \multicolumn{2}{c}{HVC} & \multicolumn{2}{c}{LVC-BC} & \multicolumn{2}{c}{LVC-NC} \\
         & $v_0 ~ [\rm{km~s^{-1}}]$ & FWHM $[\rm{km~s^{-1}}]$ & $v_0 ~ [\rm{km~s^{-1}}]$ & FWHM $[\rm{km~s^{-1}}]$ & $v_0 ~ [\rm{km~s^{-1}}]$ & FWHM $[\rm{km~s^{-1}}]$ \\
        \hline
        $[\ion{O}{I}]$ 5577 & $-$ & $-$ & $-15.7 \pm 1.5$ & $72 \pm 3$ & $-$ & $-$ \\
        $[\ion{O}{I}]$ 6300 & $-206 \pm 1$ & $97 \pm 2$ & $-55.9 \pm 2.2$ & $146 \pm 4$ & $-21.8 \pm 0.4$ & $36 \pm 1$ \\
        $[\ion{S}{ii}]$ 6731 & $-216.6 \pm 0.2$ & $68.3 \pm 0.6$ & $-79.8 \pm 0.6$ & $114 \pm 1$ & $-28.5 \pm 0.1$ & $26.9 \pm 0.3$ \\
        \hline
        \end{tabular}
        \tablefoot{{$v_0$ and FWHM are the centroid velocity and the full width at half maximum of each component, respectively.}}
        \label{tab:FELs_gaussian}
    \end{table*}
    
    We can distinguish three different components in the {forbidden lines}. The first one is the HVC that is observed in the [\ion{O}{i}] 6300 and [\ion{S}{ii}] 6731 lines. It traces a jet that extends out to $\sim 100$~AU from the star \citep{Whelan+2021}.
    The second component is the LVC-NC, observed in the same transitions as the HVC. This component is formed in the MHD wind launched from the outer disk \citep{Whelan+2021}. Both the HVC and the LVC-NC have been imaged with MUSE \citep{Birney+2024}.

    The third component is the LVC-BC, {which} is most evident in the [\ion{O}{i}] 5577 line. It is also observed in the [\ion{O}{i}] 6300 {and [\ion{S}{ii}]~6731 lines, but in [\ion{S}{ii}] it does not extend to redshifted velocities.}
    {The LVC-BC of the [\ion{O}{i}] 5577 line has a blueshifted centroid velocity, $v_0 = -15.7 \pm 1.5 ~ \rm{km~s^{-1}}$, which is compatible with formation in a wind, and a full width at half maximum (FWHM) of $72 \pm 3$~km\,s$^{-1}$. Using this value to approximate the broadening velocity, we can estimate the launching radius of the outflow \citep[e.g.][]{BanzattiPontoppidan2015, Fang+2018, Campbell-White+2023}. Assuming purely Keplerian rotation, the launching radius $r_0$ is
    \begin{equation}
        r_0 = \frac{G M_{\star}}{(\rm{FWHM} / \sin i_\star)^2}.
    \end{equation}
    Using the parameters of RU~Lup from Table~\ref{tab:stellar_pars}, we obtain $r_0 \approx 0.68~R_{\star}$. This value is significantly smaller than the expected truncation radius of the disk, $R_{\rm T} \approx 2~R_\star$ (Table~\ref{tab:stellar_pars}), indicating that the [\ion{O}{i}] 5577 emission cannot originate from a region purely dominated by Keplerian rotation. Therefore, the observed broadening must include contributions from non-Keplerian motions, such as a poloidal outflow component, thermal broadening, or MHD turbulence in an MRI-active disk. However, at temperatures $T \lesssim 10000$~K, the thermal broadening is only $\lesssim 3 ~ \rm{km~s^{-1}}$. Therefore, explaining the observed line width requires invoking MRI-driven turbulence or a more complex line formation process. Given the high critical density of the [\ion{O}{i}] 5577 line, we conclude that it is likely formed at the base of an outflow launched from the very inner disk.}
    
    {Compared to the [\ion{O}{i}] 5577 line, the [\ion{O}{i}] 6300 line has a lower critical density. The best-fit to its LVC-BC (Fig.~\ref{fig:FELs_superposed}) reveals a more blueshifted centroid and a broader profile than [\ion{O}{i}] 5577, consistent with higher poloidal and toroidal velocities. In contrast, the LVC-BC [\ion{S}{ii}] 6731 line ($n_{\rm c} = 1.6 \times 10^4 ~ \rm{cm^{-3}}$) shows only a blueshifted component without a red wing, which likely indicates emission from the outermost, low-density layers of the wind where rotation has dissipated and the flow is dominated by poloidal motion.}

    \section{Discrete absorption components}
    \label{DACs}

    \begin{figure}
        \centering
        \includegraphics[width=\linewidth]{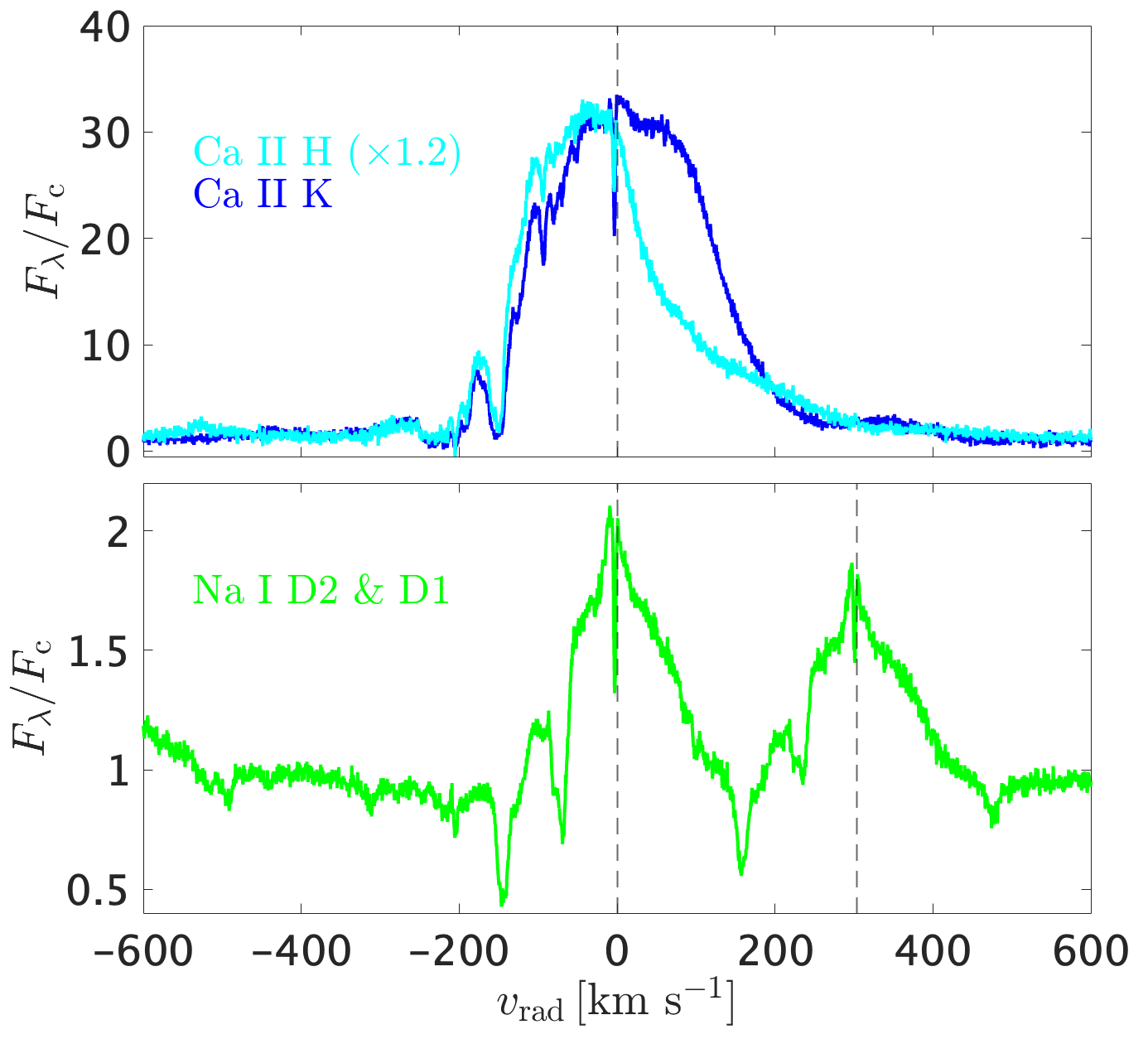}
        \caption{\ion{Ca}{ii}~H \& K and \ion{Na}{i}~D$_2$ \& D$_1$ doublets in the ES~22.5 spectrum of RU~Lup. The \ion{Ca}{ii}~H \& K lines are plotted in velocity relative to their rest wavelengths, marked with a vertical dashed line. The \ion{Na}{i} doublet is plotted in velocity relative to the rest wavelength of the D$_2$ line. The line centers of each line are indicated by the vertical dashed lines.}
        \label{fig:CaII_NaI_doublets}
    \end{figure}

    \begin{figure*}
        \centering
        \includegraphics[width=\linewidth]{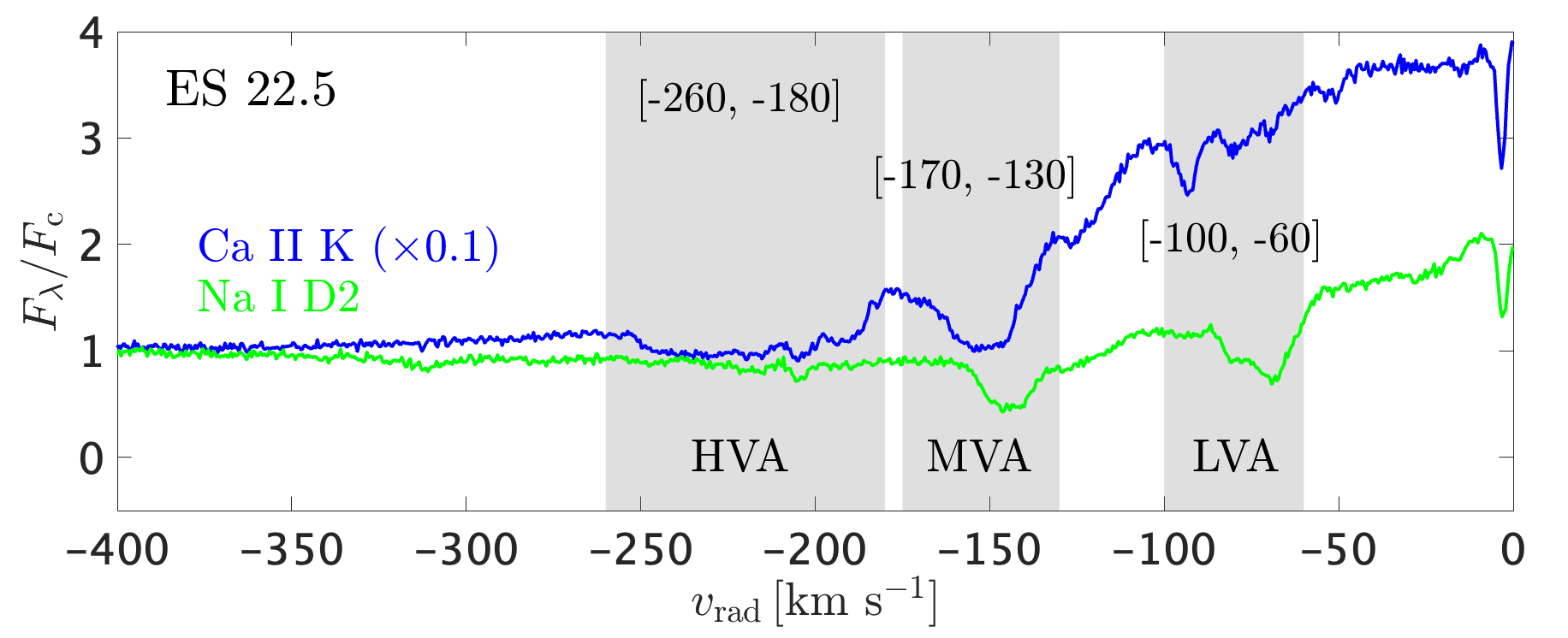}
        \caption{Discrete absorption components observed for the \ion{Na}{i}~D$_2$ (green) and \ion{Ca}{ii}~K (blue) lines in the ES~22.5 spectrum of RU~Lup. 
        The shaded areas mark the velocity ranges where absorption is observed.}
        \label{fig:CaII_vs_NaI}
    \end{figure*}
    
    RU\,Lup shows prominent emission in metallic species (Paper I). 
    In this work we focus on two strong resonance lines, the \ion{Na}{i} D$_2$ and the \ion{Ca}{ii} K lines. 
    Both lines are one of the two components of a doublet, namely the \ion{Na}{i}~D doublet (D$_1$ at $\lambda = 5895.92$~{\AA}, D$_2$ at $\lambda = 5889.95$~{\AA}) and the \ion{Ca}{ii}~H \& K doublet (H at $\lambda = 3968.47$~{\AA}, K at $\lambda = 3933.66$~{\AA}). 
    These lines have been already analyzed with UVES by \citet{Gahm+2013}, who noticed a series of absorption components in their blue wings, {at velocities between $-250 ~ \rm{km~s^{-1}}$ and $-100 ~ \rm{km~s^{-1}}$}. However, the unprecedented resolution of ESPRESSO allows a detailed study of the line profiles.
    The line profiles are shown in Fig.~\ref{fig:CaII_NaI_doublets} on the example of the ES~22.5 spectrum. 

    Figure~\ref{fig:CaII_vs_NaI} zooms in to the blue wing of the \ion{Ca}{ii}~K and \ion{Na}{i}~D$_2$ lines in the ES~22.5 spectrum. The broad ($\sim 350-400 ~ \rm{km~s^{-1}}$) emission is locally attenuated in a set of discrete absorption components. These components are not photospheric, because they are much broader than $v\sin i = 8.6 \pm 1.4 ~ \rm{km ~ s^{-1}}$ ({Table~\ref{tab:stellar_pars}}). {Therefore, they originate in the outflowing gas. The narrow absorption close to $\sim 0 ~ \rm{km ~ s^{-1}}$ is from the interstellar medium {\citep{Pascucci+2015}}.}
    {To systematize the analysis of these absorption features, we group them into three main components based on their centroid velocities in ES~22.5 and persistence across spectra. We distinguish: 
    \begin{itemize}
        \item a high velocity absorption component (HVA) that extends between $-260 ~ \rm{km ~ s^{-1}}$ and $-180 ~ \rm{km ~ s^{-1}}$,
        \item a medium velocity absorption component (MVA) that extends between $-170 ~ \rm{km ~ s^{-1}}$ and $-130 ~ \rm{km ~ s^{-1}}$,
        \item a low velocity absorption component (LVA) that extends between {$-100 ~ \rm{km ~ s^{-1}}$} and $-60 ~ \rm{km ~ s^{-1}}$. 
    \end{itemize}
    This classification is empirical and reflects the presence of preferred velocities in the absorbing gas. These velocity intervals were chosen to isolate the most prominent and recurrent absorption features, allowing for a meaningful comparison of their properties and variability.}
    
    The optical depth is different from {feature} to {feature}, and between the \ion{Ca}{ii}~K and \ion{Na}{i}~D$_2$ lines.
    The HVA component is well defined only in \ion{Ca}{ii}~K.  
    The MVA is present in both {species}, although with a different velocity structure. It is more extended in \ion{Ca}{ii}~K, and the profile minimum is more blue-shifted in \ion{Ca}{ii}~K than in \ion{Na}{i}~D$_2$. {The LVA is most clearly observed in \ion{Na}{i}~D$_2$, where it appears as a broad and structured absorption feature. A weaker and narrower absorption is also present in the same velocity range in \ion{Ca}{ii}~K, although its centroid is slightly more blueshifted.}
    These differences indicate that these {discrete features} originate in distinct regions of the outflow. While neutral sodium likely probes a colder region of the wind, absorption in \ion{Ca}{ii} requires ionized material.

    \subsection{Variability}
    \label{DACsvariability}
\begin{figure*}
        \centering
        \includegraphics[width=0.95\linewidth]{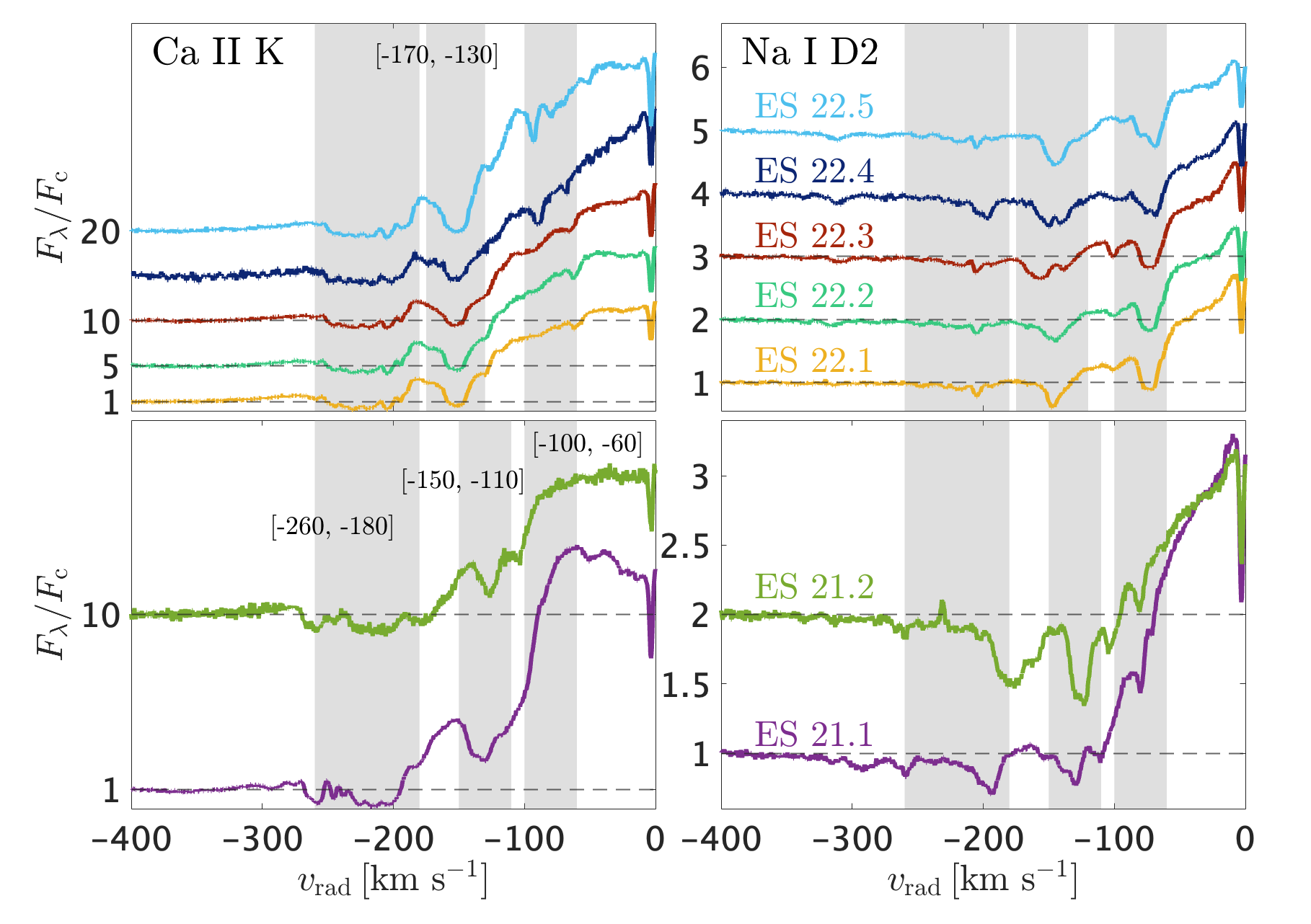}
        \caption{Variability of the \ion{Ca}{ii}~K and \ion{Na}{i}~D$_2$ lines in the high resolution ESPRESSO spectra (Table~\ref{tab:log_specobs}). The shaded areas mark the velocity ranges where absorption is observed. These ranges are the same as in Fig.~\ref{fig:CaII_vs_NaI} for the spectra from 2022, while the velocity range of the MVA is different in 2021, being $[-150, -110] ~ \rm{km~s^{-1}}$.}
        \label{fig:CaII_K_ES}
    \end{figure*}

    We used the ESPRESSO spectra to study the variability of the discrete absorption components. Figure~\ref{fig:CaII_K_ES} shows the blue wings of the \ion{Ca}{ii}~K and \ion{Na}{i}~D$_2$ lines for all existing ESPRESSO observations of RU~Lup.

    The discrete components are well defined and structured in the 2022 spectra, whereas in 2021 they appear more complex and irregular, especially in the \ion{Na}{i} D$_2$ line. {The pronounced changes in shape and depth between the two 2021 spectra indicate a more dynamic wind, coincident with an epoch of higher accretion rate \citep[Paper~I and][]{Wendeborn+2024_1}. This supports a direct link between the outflowing gas producing these features and the accretion process.}

    A key qualitative difference is in the morphology of the absorption. In 2021, the \ion{Na}{i} D$_2$ components have blue-skewed profiles, that is, single, asymmetric troughs whose depth varies monotonically across the profile. Such profiles naturally arise when the line of sight $\Delta s$ intersects an accelerating flow, so that absorption samples a monotonic velocity gradient. The characteristic broadening from such a line-of-sight gradient is $\Delta v_{\rm los} \simeq |dv/ds|\,\Delta s$. As the gas accelerates and expands, the absorption is expected to drift to higher blueshifts, consistent with the rapid, irregular variability seen in 2021.
    
    In contrast, the absorption profiles in 2022 are often Gaussian-like or double-dipped, with remarkably stable centroids within an epoch. The MVA, for instance, remains centered at approximately $-150~\mathrm{km\,s^{-1}}$ throughout 2022. At this speed the gas would travel $\sim 8\,R_\star$ per day (for $R_\star = 2.27\,R_\odot$, Table~\ref{tab:stellar_pars}). Since we observe no significant velocity drifts over several days, this implies that the absorbing gas has already reached its terminal velocity.

    While the MVA centroid velocity is stable within each observing epoch, we detect a $\sim -20~\mathrm{km\,s^{-1}}$ shift in its centroid between 2021 and 2022 in both \ion{Ca}{ii}~K and \ion{Na}{i} D$_2$. This suggests that the initial conditions of the outflow, such as the initial velocity or the launching angle, changed over time, possibly in response to the evolving accretion state of the system.

    \subsection{Comparison with forbidden emission lines}
    \label{DACs_vs_FELs}

\begin{figure}
        \centering
        \includegraphics[width=\linewidth]{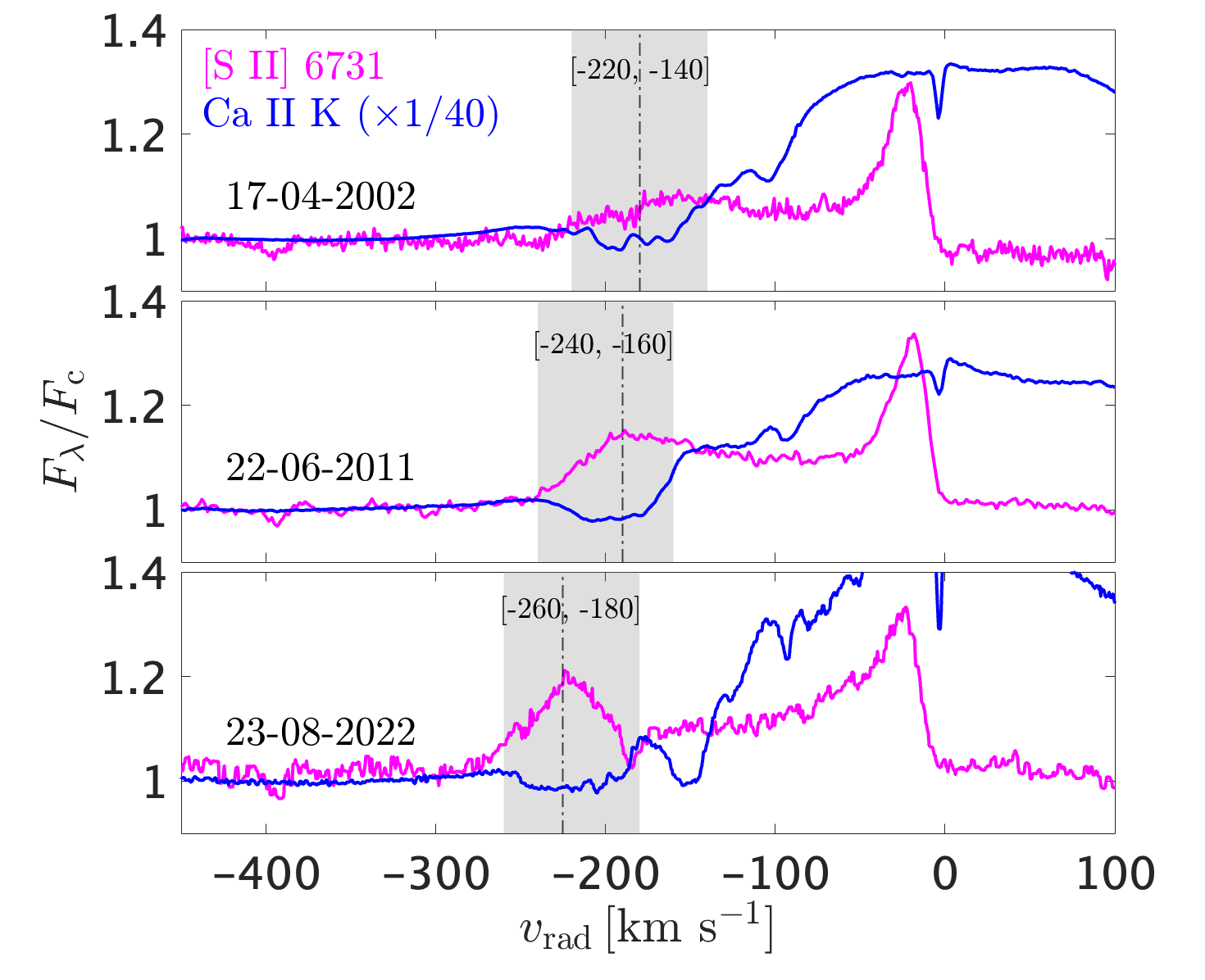}
        \caption{Comparison between the \ion{Ca}{ii}~K (blue) and [\ion{S}{ii}]~6731 (magenta) lines for a selection of spectra of RU~Lup. From the top panel to the bottom panel, the spectra are from UVES, ESPaDOnS, and ESPRESSO. The \ion{Ca}{ii}~K lines were scaled by $1/40$. The shaded areas trace the evolution of the two components, while the vertical lines mark the {peak} $v_{\rm rad}$ of the HVC.}
        \label{fig:[SII]_vs_CaII}
    \end{figure}

    The HVA observed in the \ion{Ca}{ii}~K line has the same velocity range as the HVC of the forbidden lines (see Fig.~\ref{fig:[SII]_vs_CaII}), indicating that these two features are connected.
    In Fig.~\ref{fig:[SII]_vs_CaII} we supplemented the ES~22.5 observation (bottom panel) with two archival spectra. One is from ESPaDOnS (middle panel) and it is part of the spectra analyzed by \citet{Stock+2022}. The other one is from UVES (top panel) and it is part of the observations discussed in \citet{Stempels+2007} and \citet{Gahm+2008}.
    The blueshifted extension of the [\ion{S}{ii}]~6731 line profile evolves in the same way as the \ion{Ca}{ii}~K HVA. 
    In particular, the two components drift toward higher blueshifted velocities and the HVC gets stronger relative to the continuum with time.

    The two ESPRESSO epochs from 2021 are separated by $\Delta t = 9.06$ days (Table~\ref{tab:log_specobs}). Interpreting the variability of the discrete absorption components as a disturbance propagating along the flow at a characteristic velocity $v \sim 150~\rm{km\,s^{-1}}$ (MVA in Fig.~\ref{fig:CaII_vs_NaI}), the upper limit to the distance from the launching region is $L \lesssim v\,\Delta t \approx 0.78~\rm{AU}$. Conversely, the forbidden lines in RU~Lup are stable on such short baselines \citep[e.g.,][]{Stock+2022}. Indeed, Fig.~\ref{fig:[SII]_vs_CaII} shows that the HVC evolves on decade-long timescales, consistent with an extended, slowly varying jet component.
    This suggests that, although portions of the forbidden-line profiles fall within the LVA/MVA velocity range, these two tracers are formed in different regions of the outflow. The discrete components likely probe a column of gas close to the line-emitting region, whereas the forbidden lines integrate over an extended volume, sampling different streamlines and heights in the outflow.
    
    \subsection{Velocity decomposition of the absorbing gas}
    \label{poloidal_toroidal_decomposition}

    \begin{figure}
        \centering
        \includegraphics[width=\linewidth]{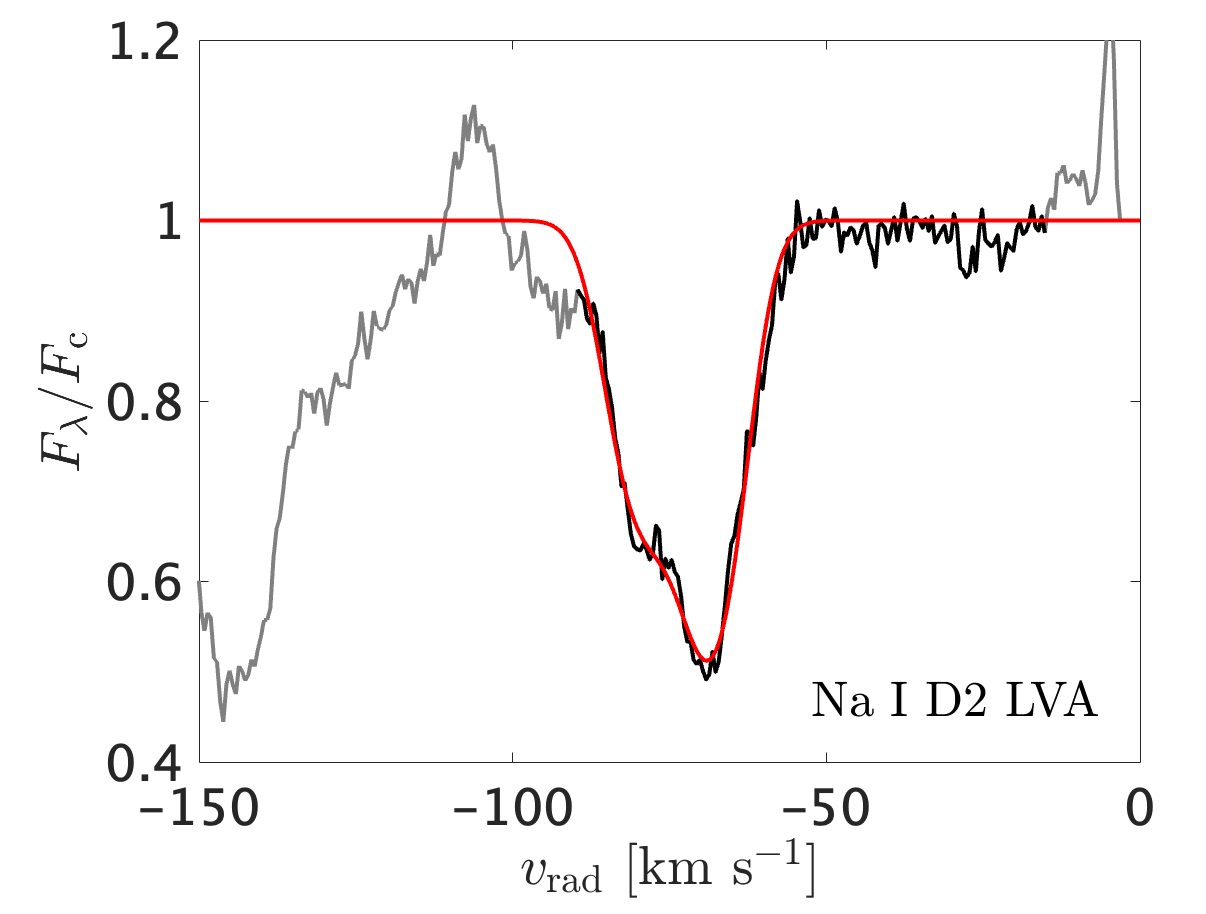}
        \caption{Best fit (red line) of the normalized absorption profile of the LVA component in the \ion{Na}{i}~D$_2$ line from the ES~22.5 spectrum (black line). The profile is the same as that of Fig.~\ref{fig:LVA_EW}. The regions excluded from the fit are plotted in gray.}
        \label{fig:LVA_best_fit}
    \end{figure}
    
    \begin{figure}
        \centering
        \includegraphics[width=\linewidth]{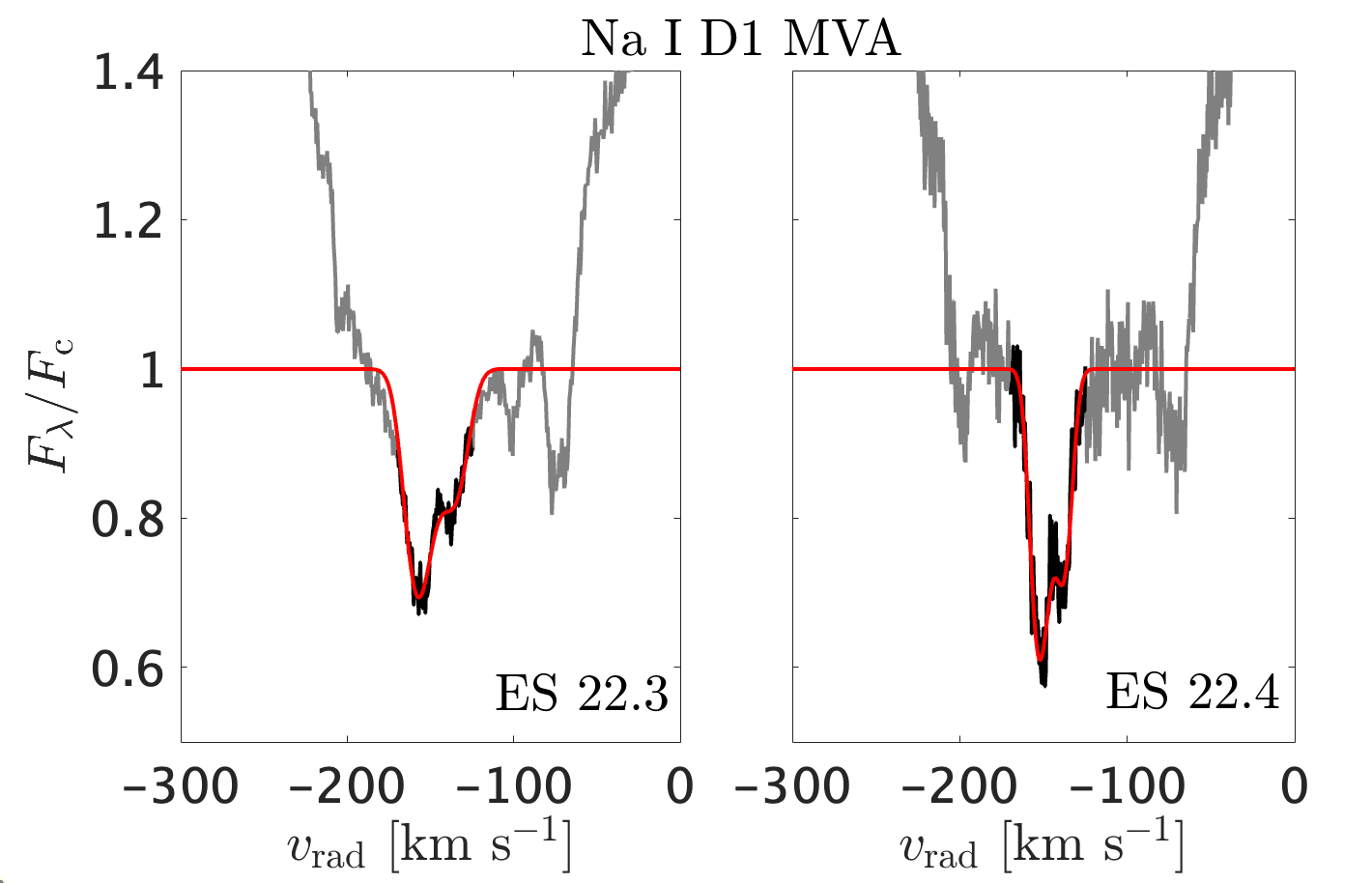}
        \caption{{Same as Fig.~\ref{fig:LVA_best_fit} but for the MVA component in the \ion{Na}{i}~D$_1$ line from the ES 22.3 and ES~22.4 spectra.}}
        \label{fig:MVA_best_fit}
    \end{figure}

    In 2022, the discrete absorption components in both \ion{Ca}{ii}~K and \ion{Na}{i}~D$_2$ are sometimes double-dipped, with stable centroids within an epoch (Sect.~\ref{DACsvariability}). Some examples are the LVA in the ES~22.5 spectrum or the MVA in the ES 22.3 and 22.4 spectra. In this section, we exploit these double-dipped profiles to measure the rotational velocity of the outflow.
    
    The velocity of the outflow can be described in cylindrical components $(v_{\rm r}, v_\phi, v_{\rm z})$, with the stellar rotation axis along $\hat{z}$. The radial velocity of the gas is $v_{\rm rad} = - v_{\rm r} \cos \phi \sin i_{\star} + v_{\phi} \sin \phi \sin i_{\star} - v_{\rm z} \cos i_{\star}$ (Appendix~\ref{ring_model}).
    At typical temperatures of the outflow ($T \lesssim 10000$~K), thermal broadening is only $\sim 1 ~ \rm{km~s^{-1}}$, far below the observed $\sim 20$-$40 ~ \rm{km\,s^{-1}}$ widths of the components. In a nearly face-on geometry, to match the full absorption width in an accelerating flow, the (vertical) velocity gradient along the line of sight should be given by $\Delta v_{\rm los} \approx |dv_{\rm z}/ds|\,\Delta s\,\cos i_\star$. 
    MHD simulations of inner winds in CTTSs have showed that the flow accelerates on a scale of a few stellar radii \citep{Romanova+2009, ZanniFerreira2013}. Using a representative value of $\Delta s = R_{\rm T} \sim 0.02$~AU, to explain a line width of $\sim 20 ~ \rm{km\,s^{-1}}$ we require $|dv_{\rm z}/ds| \sim 10^3~\rm{km\,s^{-1}\,AU^{-1}}$. Such large gradients would imply blue-skewed troughs rather than symmetric profiles, and measurable centroid drifts as the sampled column moves along the line of sight. Indeed, the expected acceleration is
    \begin{equation}
    \frac{dv_{\rm los}}{dt} \approx \frac{dv_{\rm z}}{ds} \frac{ds}{dt} \, \cos i_\star \approx  \frac{dv_{\rm z}}{ds} v_{\rm z} \cos i_\star.
    \end{equation}
    For $v_{\rm z} = 150~\rm{km\,s^{-1}}$, $i_\star = 16^\circ$, and the previously computed $|dv_{\rm z}/ds|$, this predicts a drift of $dv_{\rm los}/ dt \approx 80 ~\rm{km\,s^{-1}\,day^{-1}}$, which is not observed in 2022. Therefore, we conclude that a velocity gradient along the line of sight cannot be the dominant broadening agent in the absorption components in 2022.

    On the other hand, any finite azimuthal coverage $\Delta\phi$ at roughly fixed cylindrical radius produces a spread in $v_{\rm rad}$ of $\Delta v_{\rm rot} \simeq 2v_\phi \sin i_\star\, \sin (\Delta\phi/2)$. 
    With $i_\star=16^\circ$ and a moderate azimuthal extent $\Delta\phi = \pi/2$, we find $\Delta v_{\rm rot} \approx 0.4\,v_\phi$. Therefore, toroidal velocities of $\sim 50~\rm{km\,s^{-1}}$ are compatible with the observed line widths and naturally explain the stable centroids. Changes in $v_\phi$ and $\Delta\phi$ modulate the width, while the centroid remains set by $-v_z\cos i_\star$.

    Motivated by the above kinematic considerations, we modeled each discrete component as absorption from a sector of a narrow ring at cylindrical position $(r,z)$ above the disk. Although the absorber is compact along the line of sight, it can subtend a finite azimuthal range $\phi\in[\phi_1,\phi_2]$. The resulting distribution of projected velocities can produce the Gaussian-like and double-dipped profiles observed in 2022. The assumption of a ring-like geometry is supported by MHD simulations, which show that disk winds tend to exhibit cylindrical symmetry even in fully three-dimensional setups \citep{Romanova+2009}, and by the discrete nature of the observed components, which suggests localized ejection events rather than a continuous wind.
    
    In the model, the gas is assumed to be rotating and outflowing, with a velocity vector decomposed into vertical ($v_{\rm z}$) and toroidal ($v_\phi$) components. We assumed that the flow is predominantly in the vertical direction, so that $\vec{v_{\rm p}} \approx v_{\rm z} \hat{\vec{z}}$. The ring is limited between $\phi_1$ and $\phi_2$ in azimuth. The model has $7$ parameters: $v_{\rm z}$ and $v_{\phi}$, the angles $\phi_1$ and $\phi_2$, and the scaling parameters $\sigma$, $\tau_0$, and CF. A full description of the model parameters and the profile computation is given in Appendix~\ref{ring_model}.
    
    We used our model to fit the LVA of the ES~22.5 spectrum. To this end, we created a normalized profile by using the red wing of the \ion{Na}{i}~D$_2$ line as model for the unabsorbed emission, as explained in Appendix~\ref{DACs_EW}. We considered only the interval between $-90 ~ \rm{km~s^{-1}}$ and $-15 ~ \rm{km~s^{-1}}$ in the profile of Fig.~\ref{fig:LVA_EW}. Since there is a degeneracy between $\tau_0$ and CF (Fig.~\ref{fig:jet_abs_profiles}), we fixed $\rm{CF} = 1$, that is, we assumed complete coverage of the emitting region.
    The result of the fit is shown in Fig.~\ref{fig:LVA_best_fit}. The best fit has a reduced chi-square ($\chi^2_{\rm red}$) of $1.5$. The parameters are $v_z = 77.0 \pm 0.6 ~ \rm{km~s^{-1}}$, $v_{\phi} = 29.2 \pm 1.2 ~ \rm{km~s^{-1}}$, $\phi_1 = -100 \pm 21 ~ ^{\rm o}$, $\phi_2 = 163 \pm 43 ~^{\rm o}$, $\tau_0 = 0.126 \pm 0.001$, and $\sigma = 4.39 \pm 0.03 ~ \rm{km~s^{-1}}$. The best-fit azimuthal limits suggest that the region is significantly extended in azimuth, with $\Delta \phi = |\phi_2 - \phi_1| \approx 3\pi/2$.
    
    A similar analysis was performed on the MVA component in the ES~22.3 and 22.4 spectra. In this case, we used the \ion{Na}{i}~D$_1$ line instead of the D$_2$. This is because the MVA of the D$_1$ line is between the \ion{Na}{i}~D$_2$ redshifted wing and the \ion{Na}{i}~D$_1$ blueshifted wing, making it simple to find a local continuum. Figure~\ref{fig:MVA_best_fit} shows the result of the fits. 
    For the MVA in ES~22.3, the best fit yields $v_{\rm z} = 152.3 \pm 1.6 ~ \rm{km~s^{-1}}$, $v_{\phi} = 55.2 \pm 1.3 ~ \rm{km~s^{-1}}$, $\phi_1 = -160 \pm 35 ~^{\circ}$, $\phi_2 = 90 \pm 20 ~^{\circ}$, $\tau_0 = 0.112 \pm 0.002$, and $\sigma = 6.74 \pm 0.04 ~ \rm{km~s^{-1}}$, with $\chi^2_{\rm red} = 1.19$.
    For the MVA in ES~22.4, the best fit parameters are $v_z = 151.5 \pm 0.4 ~ \rm{km~s^{-1}}$, $v_{\phi} = 34.8 \pm 1.8 ~ \rm{km~s^{-1}}$, $\phi_1 = -165 \pm 33 ~ ^{\circ}$, $\phi_2 = 113 \pm 15 ~^{\circ}$, $\tau_0 = 0.106 \pm 0.002$, and $\sigma = 4.45 \pm 0.04 ~ \rm{km~s^{-1}}$, with $\chi^2_{\rm red} = 1.94$. 
    The remaining components were not modeled, as they do not exhibit the distinctive double-dipped morphology necessary for an application of our fitting procedure.

    Our ring model constrains the projected rotation $V \equiv v_{\phi}\sin i_\star$, so the uncertainty on $i_\star$ propagates directly into the inferred $v_{\phi}$. In our fits we fix $i_\star$ to its central value ($16^\circ$; Table~\ref{tab:stellar_pars}) but holding $V$ constant, differentiation gives $\sigma_{v_{\phi}}/v_{\phi} \approx \sigma_{i_\star}/\tan i_\star$, with $\sigma_{i_\star}$ in radians. At $i_\star=16^\circ$ and $\sigma_{i_\star}=5^\circ$ (Table~\ref{tab:stellar_pars}), this corresponds to a relative uncertainty of $\sim 30\%$ on $v_{\phi}$. For example, for the LVA in ES~22.5 we obtain $\sigma_{v_{\phi}}\approx 8.9~\mathrm{km\,s^{-1}}$. This contribution of the inclination to the uncertainty in $v_{z}$ is comparable to the uncertainties of the formal fit. Therefore, we neglect it and quote the fitted $\sigma_{v_{z}}$. Table~\ref{tab:LVA_MVA_summary} reports the values of $v_{\rm z}$ and $v_\phi$ derived for the LVA and MVA. 
    
    \subsection{Launching radii and angular momentum removal} 
    \label{DACs_properties}
   
    The distinct properties of the LVA and MVA components suggest that they originate in physically different regions of the outflow. The discrete nature of the components argues against an origin in a stellar wind. Such winds are expected to produce broad absorption profiles, like those observed in the \ion{He}{i} 10830 line \citep[e.g.,][]{Edwards+2006, Erkal+2022}. In contrast, it suggests a confined and transient origin, consistent with ejections from the disk.
    
    In the cold disk wind framework, the vertical velocity $v_{\rm z}$ and the specific angular momentum $\ell = r v_\phi$ depend on the launching radius $r_0$ and the magnetic lever arm $\lambda$, defined as $\lambda = (r_{\rm A} / r_0)^2$. Here $r_{\rm A}$ is the Alfvén radius, that is, the point at which the poloidal velocity equals the Alfvén speed, $v_{\rm A} = B / \sqrt{4\pi \rho}$. The lever arm describes the efficiency of angular momentum extraction. Conservation of angular momentum beyond the Alfvén radius gives
    \begin{equation}
        \ell = r\,v_\phi = \lambda\,\sqrt{G M_\star r_0}.
        \label{eq:ell_cold_disk_theory}
    \end{equation}
    The terminal $v_{\rm z}$ is linked to the Keplerian velocity at $r_0$, $v_{\rm K}(r_0) = \sqrt{GM_{\star}/r_0}$, that is,
    \begin{equation}
        v_{\rm z} \simeq \sqrt{2\lambda - 3} \; \sqrt{\frac{G M_\star}{r_0}}
        \label{eq:vz_cold_disk_theory}
    \end{equation}
    \citep[e.g.,][]{Ferreira+2006}.
    In this framework, different values of $v_{\rm z}$ suggest different launching properties.
    
    Unlike forbidden emission lines imaged at high spatial resolution, the absorption features are not spatially resolved. This means that the cylindrical radius of the absorbing gas, $r_{\rm abs}$, hence $\ell = r_{\rm abs}v_{\phi}$, cannot be directly determined. {However, $r_{\rm abs}$ can be linked to the launching radius by calculating where the observer’s line of sight to the magnetosphere intersects a streamline that originates at $r_0$ in the disk plane. Assuming that the streamline has an inclination $\theta_{\rm s}$ and that the emission comes from a cylindrical radius $r_{\rm mag}$ in the disk plane, the intersection yields
    \begin{equation}
        r_{\rm abs} = \frac{ r_0 \tan i_\star + r_{\rm mag}\,\tan\theta_s }{ \tan i_\star - \tan\theta_s} \equiv Kr_0 + \alpha r_{\rm mag}
        \label{eq:rabs_full}
    \end{equation}
    (Appendix~\ref{rabs_derivation}).
    Together with the cold disk wind equations (Eqs.~\eqref{eq:ell_cold_disk_theory} and \eqref{eq:vz_cold_disk_theory}), this equation allows the determination of $r_0$, $\lambda$, and $r_{\rm abs}$ as a function of the measured velocities $v_{\rm z}$ and $v_{\phi}$ and the parameters $i_{\star}$, $\theta_{\rm s}$, and $r_{\rm mag}$. In particular, introducing $y = \sqrt{r_0}$, we can obtain $r_0$ from the cubic equation
    \begin{equation}
        \frac{v_z^2}{G M_\star}\,y^3
        - \frac{2 K\,v_\phi}{\sqrt{G M_\star}}\,y^2
        + 3\,y
        - \frac{2 \alpha\, v_\phi\, r_{\rm mag}}{\sqrt{G M_\star}}
        = 0.
    \label{eq:cubic}
    \end{equation}
    Details of the derivation are discussed in Appendix~\ref{r0_lambda_derivation}.}

    We studied the families of solutions by plotting $r_0$ as a function of $r_{\rm mag}$ for a grid of $\theta_{\rm s}$ values at fixed $i_\star=16^\circ$. The details are reported in Appendix~\ref{r0_vs_rmag}. The key outcome is that imposing an upper bound on $r_{\rm abs}$ ($\lesssim L\sin i_\star$ with $L=1$~AU) limits the launching radius to $r_0 \lesssim 6.76\,R_\star$ for both components. Then, using the measured $v_{\rm z}$ and the cold disk wind relations yields $\lambda_{\rm LVA}\in[1.56,\,1.93]$ and $\lambda_{\rm MVA}\in[1.75,\,3.20]$, with $\lambda_{\rm LVA}<\lambda_{\rm MVA}$ at a given $r_0$. An upper bound on the specific angular momentum carried by each component follows from the upper limit on $r_{\rm abs}$ and the measured $v_\phi$. Using the central values of $v_{\phi}$, we obtain $\ell_{\rm LVA}\lesssim 8.1~\mathrm{AU\,km\,s^{-1}}$, $\ell_{\rm MVA(ES~22.3)}\lesssim 15.2~\mathrm{AU\,km\,s^{-1}}$, and $\ell_{\rm MVA(ES~22.4)}\lesssim 9.6~\mathrm{AU\,km\,s^{-1}}$.

    \subsection{Rate of mass loss}
    \label{DACs_mass_loss}
    To investigate the contribution of the different components of the outflow to the mass and angular momentum budget of the system, we estimated the mass-loss rate associated with the LVA and MVA. The total mass-loss rate can be expressed as
    \begin{equation}
        \dot{M}_{\rm wind} = \mu m_{\rm H} \, n_{\rm H} \, A \, v_{\rm z},
    \end{equation}
    where $\mu \approx 1.3$ is the mean molecular weight for a gas of solar composition, $m_{\rm H}$ is the mass of the hydrogen atom, $n_{\rm H}$ is the hydrogen number density, $A$ is the area through which the gas flows, and $v_{\rm z}$ is the vertical velocity of the absorbing material.
    The fact that the features are observed in absorption allows us to simplify this expression. The hydrogen number density is related to the column density via $n_{\rm H} = N_{\rm H} / \Delta s$, where $\Delta s$ is the extent of the absorbing gas along the line of sight. The area $A$ can be computed by assuming the same ring-like geometry described in Sect.~\ref{poloidal_toroidal_decomposition}. The ring is located at the absorbing radius $r_{\rm abs}$, has a radial thickness $\Delta r$, and an azimuthal extent $\Delta\phi = |\phi_2 - \phi_1|$. Therefore, the area of the absorbing region is $A = \Delta\phi \, r_{\rm abs} \, \Delta r$.
    The radial thickness is related to the line-of-sight extent by $\Delta r = \Delta s \sin i_\star$, where $i_\star$ is the system inclination (see Fig.~\ref{fig:ring_abs_cylindrical}). Substituting into the previous expressions, we obtain
    \begin{equation}
        \dot{M}_{\rm wind} = \mu m_{\rm H} \, N_{\rm H} \, \Delta\phi \, r_{\rm abs} \, \sin i_\star \, v_{\rm z}.
    \end{equation}
    For both components we measured an azimuthal span $\Delta\phi \approx 3\pi/2$ (Sect.~\ref{poloidal_toroidal_decomposition}), and we adopt the conservative upper limit $r_{\rm abs} \lesssim 0.276~\mathrm{AU}$ (Appendix~\ref{r0_vs_rmag}). Therefore, the mass-loss rate of the two components differs because of their measured $N_{\rm H}$ and $v_{\rm z}$. Our estimate of $N_{\rm H}$ is derived from the equivalent width of the \ion{Na}{I}~D lines, as described in Appendix~\ref{DACs_EW}.
    
    For the LVA, the absence of \ion{Ca}{ii} absorption suggests that the absorbing gas is predominantly neutral, so that $N_{\rm H}$ follows directly from \ion{Na}{i} under the assumption of solar abundances. We obtained $N_{\rm H} = 5.49\times10^{17} ~ \rm{cm^{-2}}$ (Appendix~\ref{DACs_EW}) and $v_{\rm z}=77.0 ~ \rm{km\,s^{-1}}$ (Sect.~\ref{poloidal_toroidal_decomposition}). This translates into $\dot{M}_{\rm LVA} \lesssim 7.4 \times 10^{-13} ~ M_\odot ~ \rm{yr^{-1}}$.

    In the case of the MVA, the strength of the MVA absorption in \ion{Ca}{ii} suggests that sodium in the absorbing gas is predominantly ionized, so $N_{\rm H}$ must be inferred by correcting \ion{Na}{i} for excitation and ionization effects. In Appendix~\ref{DACs_EW} we estimated $N_{\rm H} = 1.16 \times 10^{18}\,\chi^{-1} ~ \rm{cm^{-2}}$, where $\chi = N_{\rm Na\,I}/N_{\rm Na}$ is the fraction of sodium atoms in the neutral state. Using $v_{\rm z} = 151.5~\mathrm{km\,s^{-1}}$ (Sect.~\ref{poloidal_toroidal_decomposition}), we find $\dot{M}_{\rm MVA} \lesssim 3.3 \times 10^{-12} \,\chi^{-1} ~ M_\odot~\rm{yr^{-1}}$.
    
\begin{figure}
        \centering
        \includegraphics[width=\linewidth]{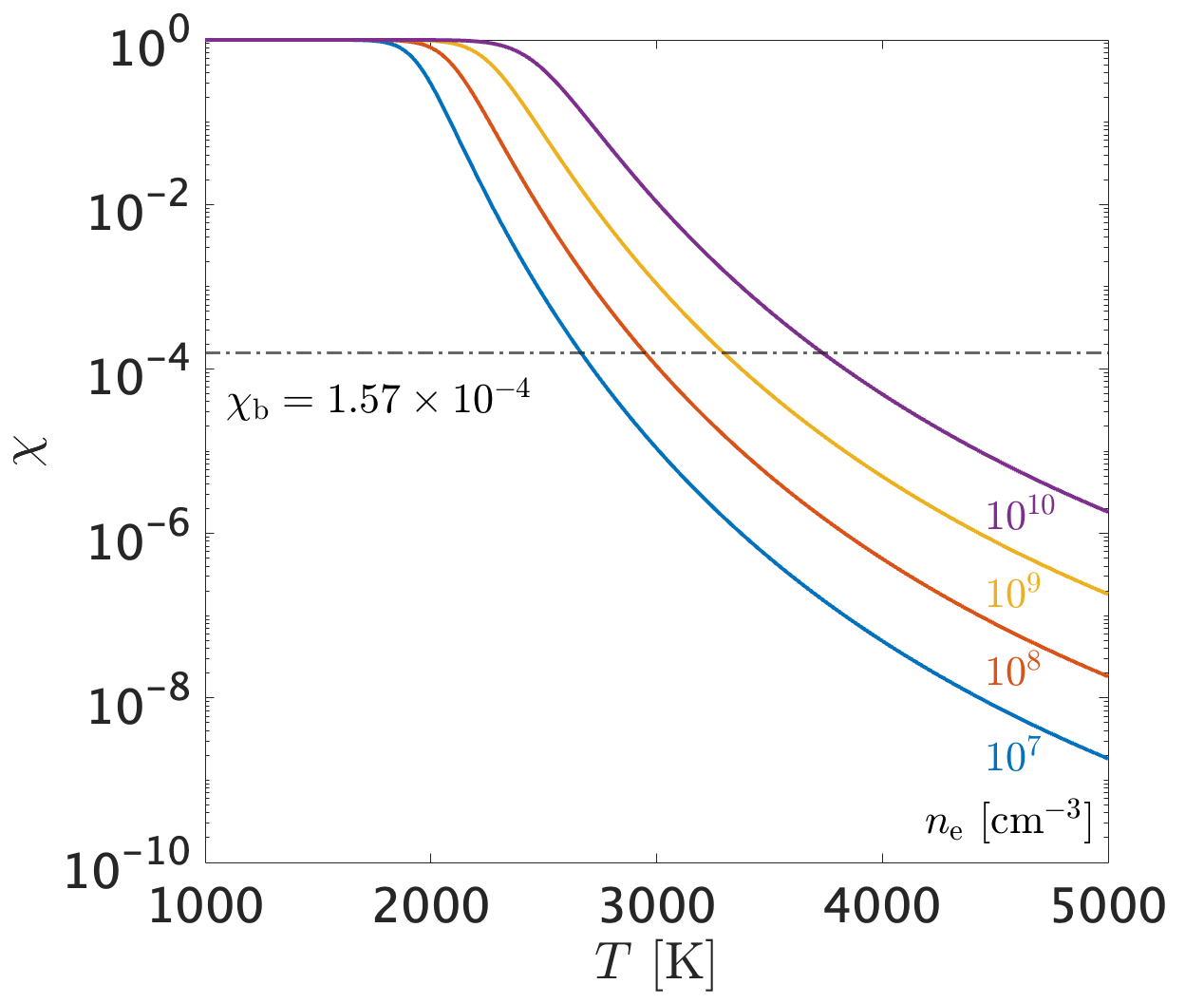}
        \caption{Neutral fraction of sodium ($\chi$) as a function of $T$, computed using the Saha equation for different values of $n_{\rm e}$. The dash-dotted line marks the critical neutral fraction $\chi_{\rm b}$ required for the MVA wind torque to balance the accretion torque (see Sect.~\ref{MEs_efficiency}).}
        \label{fig:chi_Saha}
    \end{figure}

    The ionization level of the MVA is further supported by the values of $\chi$ predicted by the Saha equation (Fig.~\ref{fig:chi_Saha}). For plausible temperatures {reached in the inner disk close to the star ($T \sim 3000$~K)}, the neutral fraction of sodium {is in the range $10^{-3}-10^{-5}$, implying that the measured \ion{Na}{i} absorption likely traces only a small fraction of the total gas column. \citet{Nisini+2018} found that in CTTSs the forbidden lines trace winds with $\dot{M}_{\rm wind} \approx 0.1 \dot{M}_{\rm acc}$. If the discrete absorption components probe similar mass-loss channels, adopting $\dot{M}_{\rm acc}\!\approx\!10^{-7}\,M_\odot\,\mathrm{yr^{-1}}$ for RU~Lup implies a neutral sodium fraction of $\chi \sim 3.3 \times 10^{-4}$.}

    \section{Discussion}
    \label{discussion}

    \begin{table*}[t]
        \centering
        \caption{{Summary of derived properties for the LVA and MVA components.}}
        \begin{tabular}{lccccc}
        \hline
        Component & $v_z$ & $v_{\phi}$ & $\ell$ & $\lambda$ & $\dot{M}_{\rm wind}$ \\
         & $[\rm km~s^{-1}]$ & $[\rm km~s^{-1}]$ & $[\rm AU~km~s^{-1}]$ & & $[M_{\odot}~\rm{yr}^{-1}]$ \\
        \hline
        LVA (ES~22.5) & $77.0 \pm 0.6$ & {$29.2 \pm 8.9$} &  {$\lesssim 8.1$} &  {$1.56-1.93$} &  {$\lesssim 7.4 \times 10^{-13}$} \\
        MVA (ES 22.3) & $152.3 \pm 1.6$ &  {$55.2 \pm 16.8$} &  {$\lesssim 15.2$} &  {\multirow{2}{*}{$1.75-3.20$}}  &  {\multirow{2}{*}{$(3.3 \times 10^{-12}) \cdot \chi^{-1}$}} \\
        MVA (ES 22.4) & $151.5 \pm 0.4$ &  {$34.8 \pm 10.6$} &  {$\lesssim 9.6$} & & \\
        \hline
        \end{tabular}
        \tablefoot{The parameter $\chi$ represents the fraction of neutral sodium in the absorbing gas (see Appendix~\ref{DACs_EW}).}
        \label{tab:LVA_MVA_summary}
    \end{table*}

    The discrete absorption components and their connection with the {forbidden lines} provide important insights into the structure of the outflow of RU~Lup. Here we outline the implications of our observations on the physics of the MHD-driven wind of RU~Lup and, by extension, of CTTSs in general. Table~\ref{tab:LVA_MVA_summary} summarizes the properties derived for the LVA and MVA.

    A variety of MHD simulations have shown that coexisting winds can be launched from different regions of the star–disk system \citep[e.g.,][]{Ferreira+2006}. Extended magneto-centrifugal disk winds, driven from the disk surface over a range of radii, naturally produce an “onion-like” flow in which the inner streamlines have higher poloidal velocity. At the same time, magnetospheric wind models \citep[e.g.,][]{Romanova+2009, KurosawaRomanova2012} suggest that matter is expelled from the star–disk boundary near the truncation radius. 
    Moreover, any stellar wind accelerated along open stellar field lines might produce a high velocity component \citep[e.g.][]{MattPudritz2005b}.
    Overall, simulations predict the coexistence of a disk wind, a magnetospheric wind, and a stellar wind. Each mechanism dominates a different region of the outflow, producing a multi-layered structure which has been confirmed by observations \citep[e.g.,][]{Bacciotti+2002, Pyo+2006, Krist+2008, Bacciotti+2025}.
    
    \subsection{Origin of the broad \ion{Ca}{ii} emission}
    \label{CaII_emission}

    In Paper\,I we showed that the rich emission spectrum of RU~Lup is formed in the accretion flow that leaves the disk at $R_{\rm T}$. The \ion{Na}{i}~D lines have excitation conditions and line profiles similar to those of the neutral metallic lines which we studied in Paper\,I, for example the \ion{Fe}{i}~5447 line.
    Hence, they must be formed in a similar way, as we have already discussed in Paper\,I. 
    
    On the other hand, Fig.~\ref{fig:CaII_vs_NaI} shows that the \ion{Ca}{ii}\,H \& K lines are much broader than the \ion{Na}{i}\,D lines, with detectable emission up to $\sim -400 ~ \rm{km~s^{-1}}$.
    \citet{Coffey+2007} observed the jet of the CTTS DG~Tau with the \textit{Hubble Space Telescope Imaging Spectrograph} (HST/STIS) in the near-ultraviolet (NUV) by placing the slit at a distance of $0.3^{\prime\prime}$ from the star, corresponding to a de-projected distance along the jet of $68$~AU. They demonstrated that the blue wing emission of the \ion{Mg}{ii}~h~\&~k doublet in DG~Tau forms within the jet. 
    Comparing their observation with an HST spectrum of DG~Tau with a $2^{\prime\prime}$ coverage centered on the star \citep{Ardila+2002} in which the \ion{Mg}{ii}~k line showed a deep blueshifted absorption, \citet{Coffey+2007} concluded that the jet is both absorbing and emitting in \ion{Mg}{ii} (Fig.~9 of their paper).
    The \ion{Mg}{ii}~h~\&~k doublet is analogous to the \ion{Ca}{ii} H \& K doublet (hence the same denomination).  
    This similarity suggests that the wings of the \ion{Ca}{ii} H \& K doublet lines also originate in the jet, and that the absorption components observed in the \ion{Ca}{ii}~K line are a manifestation of the jet {self-absoprtion}. 
    
    \subsection{HVA component: jet knot at large scale}
    \label{HVA}
    The similar velocity structure between the \ion{Ca}{ii} HVA and the HVC of the {forbidden lines} (see Fig.~\ref{fig:[SII]_vs_CaII}) indicates a common origin. This has been already observed for the CTTS V1331~Cyg by \citet{Petrov+2014}, who called the {discrete absorption components} “shell profiles”.

    For RU~Lup, \citet{Whelan+2021} showed that the HVC component of the [\ion{S}{ii}]~6371 line is formed in a “knot” of the outflow at $\sim 55$~AU along the outflow position angle (PA). The knot is likely formed due to a shock within the jet, in which ionized atoms recombine in the post-shock region \citep[e.g.,][]{Hartigan+1995}.
    The observed velocity sub-structure in the \ion{Ca}{ii} HVA is compatible with a formation in a shocked region, where the flow is likely turbulent and parcels of gas might have a local velocity component relative to the bulk velocity of the knot. 
    
    Unlike the LVA and the MVA, which trace the outflow close to the launching region, the HVA is associated with the large scale structure of the jet, that has low density and is also observed in the forbidden lines. For the HVA, the line width is likely not representative of the toroidal velocity in the flow. The lack of an HVA in \ion{Na}{i} indicates that sodium is fully ionized. Similarly, hydrogen is expected to be ionized, implying $n_{\rm H} \approx n_{\rm e}$.
    Given the lower density of the outflow, the column required to produce such an optically thick absorption component must be significantly greater than in the regions responsible for the other two absorption features. Consequently, the observed velocity range could result from the gradient in $v_z$ along the line of sight, or from turbulent broadening within the shocked region.

    \subsection{LVC-BC: conical wind}
    \label{LVC-BC}
    The HVC and LVC-NC components of the forbidden lines have a known origin \citep[][and Sect.~\ref{FELs}]{Whelan+2021}. Here we focus on the newly identified LVC-BC, which is most evident in the [\ion{O}{i}]~5577 line. This component is blueshifted and broad, with a profile that extends to redshifted velocities. These properties suggest the presence of rotation and outflowing motion in a relatively compact region. {Using the FWHM of the line, we suggested that the outflow is launched from the inner disk,} close to the disk truncation radius, $R_{\rm T} \approx 2 ~ R_{\star} = 0.02$~AU (Table~\ref{tab:stellar_pars}).
   
    The LVC-BC is also detected in the [\ion{O}{i}] 6300 and [\ion{S}{ii}] 6731 lines, and their profiles suggest a layered wind structure. The [\ion{O}{i}] 6300 component is broader and more blueshifted than [\ion{O}{i}] 5577, indicating that it traces material that has reached higher poloidal velocities and developed a stronger toroidal component farther out in the wind. In contrast, the [\ion{S}{ii}] 6731 line, which has a much lower critical density, shows only a narrow, blueshifted component without a red wing, pointing to emission from the outermost regions of the wind, where rotation has largely dissipated and poloidal motion dominates. The three lines reveal a stratified MHD wind: as gas is launched and accelerates away from the disk, it becomes progressively less dense, gains a toroidal velocity component, and ultimately reaches a regime dominated by poloidal flow.
    
    The properties of the LVC-BC fit the description of the conical wind observed in the MHD simulations {performed by} \citet{Romanova+2009}. This type of outflow is launched from a narrow region of the inner disk between $R_{\rm T}$ and $R_{\rm co}$, where the stellar magnetosphere interacts with the inner disk. 
    
    \subsection{LVA and MVA components: ejections from the inner disk}
    \label{LVA_MVA_discussion}

    {The kinematic constraints derived in Sect.~\ref{DACs_properties} and Appendix~\ref{r0_vs_rmag} point to a picture in which the LVA and MVA trace outflows launched from the inner disk, at $r_0 \lesssim 6.76\,R_\star$. Using the measured $v_z$ in the cold disk wind approximation (Eq.~\eqref{eq:vz_cold_disk_theory}) yields $\lambda$ intervals of $\lambda_{\rm LVA}\in[1.56,\,1.93]$ and $\lambda_{\rm MVA}\in[1.75,\,3.20]$, where the lower limits are obtained assuming $r_0 \geq R_{\star}$. The geometry confines the streamline inclination to $\theta_{\rm s}\sim 12^\circ$--$14^\circ$.} 

    Despite similar kinematic properties, the two discrete absorption components appear to have different ionization stages. The LVA is detected only in \ion{Na}{i}, whereas the MVA is seen in both \ion{Na}{i} and \ion{Ca}{ii}. After correcting the column density of \ion{Na}{i} for ionization, this would suggest that the MVA carries a larger mass flux. To interpret this in the framework of magneto–centrifugal disk winds, we introduce the ejection efficiency 
    \begin{equation}
         \xi = \frac{{\rm d}\ln \dot{M}_{\rm acc}}{{\rm d}\ln r},
    \end{equation}
    which measures the fractional mass loading of the wind per logarithmic interval in launching radius \citep{CasseFerreira2000A}. In the cold disk wind theory, the magnetic lever arm is linked to $\xi$ by $\lambda \simeq 1 - 1/2\xi$ \citep{Ferreira1997}, so that similar lever arms imply similar ejection efficiencies. The total wind mass-loss rate from an annulus $[r_{\rm in}, r_{\rm out}]$ is
    \begin{equation}
        \dot{M}_{\rm wind} \;\simeq\; \xi\,\dot{M}_{\rm acc}(r_{\rm in})\,\ln\!\left(\frac{r_{\rm out}}{r_{\rm in}}\right)
        \label{eq:wind_efficiency}
    \end{equation}
    \citep[e.g.,][]{CasseFerreira2000A}. With similar values of $\xi$, a three-order-of-magnitude difference in $\dot{M}_{\rm wind}$ cannot be explained by a plausible difference in $\ln(r_{\rm out}/r_{\rm in})$ alone. This indicates that the LVA and MVA have comparable mass fluxes, hence similar ionization levels. 
    In this scenario, the non-detection of the LVA in \ion{Ca}{ii} can be explained if the LVA covers a very small portion of the \ion{Ca}{ii} emitting background, underestimating the ionization level and thus $\dot{M}_{\rm wind}$. This is plausible if the low-velocity portion of the \ion{Ca}{ii} emission does not originate primarily in the magnetosphere, but rather in a region of the inner wind closer to the axis, as we suggested in Sect.~\ref{CaII_emission}. We quantify this scenario in Appendix~\ref{MVA_LVA_coverage}, where we show that under the assumption that the \ion{Ca}{ii} emission arises in an inner, axial outflow, the absorbing gas can be arranged in two broader shells. The MVA is just outside the emitting region and the LVA further out. In this picture, our line of sight to the emitting region traverses the MVA layer, yielding a large line–emitter covering fraction but a small continuum covering and producing deep absorption that remains above the continuum. Conversely, the LVA crosses the continuum emission but largely misses the \ion{Ca}{ii} emitter, so at the LVA velocity the absorption is diluted to a few-percent level.
    
    The ejection index $\xi$ measures the local efficiency of mass loss from the accretion disk into the wind, reflecting how rapidly $\dot{M}_{\rm acc}$ decreases with $r$. From the measured $\lambda$, we infer $\xi_{\rm LVA} \approx 0.54\text{--}0.89$ and $\xi_{\rm MVA} \approx 0.23\text{--}0.67$. These values point to “warm” MHD winds characterized by efficient mass loading \citep{CasseFerreira2000B}. Such high values of $\xi$ exceed those typical of cold MHD winds, which usually exhibit $\xi \sim 0.01$ \citep{Ferreira1997}, that is, a minimal mass loading. Warm disk wind models invoke surface heating mechanisms that increase the gas thermal energy and pressure scale height, enhancing the vertical pressure gradient at the disk surface. This yields stronger mass ejection, leading to higher $\xi$ values in the range $0.1 - 0.3$ or more, depending on heating strength \citep{CasseFerreira2000B,FerreiraCasse2004}. Our derived values suggest winds with significant thermal energy contributions, consistent with heated disk atmospheres or episodic ejections from the magnetosphere, as discussed in MHD models \citep{Ferreira+2006,ZanniFerreira2013}. 
    
    The MVA exhibits a vertical velocity about a factor of two larger than the LVA. If $\lambda$ were the same for both components, a purely inward shift of the footpoint would require $r_0$ to decrease by a factor of four. Assuming a launching for the LVA at our inferred upper limit ($r_0 = 6.76 ~ R_{\star}$) would yield $r_0 = 1.69 ~ R_{\star}$ for the MVA, that is, too close to the stellar surface. Thus, launching closer to the star cannot fully account for the observed $v_{\rm z}$ of the MVA, suggesting that the MVA either traces streamlines with a larger lever arm $\lambda$ or has a non-negligible enthalpy at its base. In the second case Eq.~\eqref{eq:vz_cold_disk_theory} is modified to
    \begin{equation}
       v_z \;\simeq\; \sqrt{\,2\lambda - 3 + \beta\,}\;\sqrt{\frac{G M_\star}{r_0}},
        \label{eq:vz_warm_disk_wind}
    \end{equation}
    where $\beta$ parametrizes the extra enthalpy available to accelerate the gas \citep{Ferreira+2006}. Assuming a nearly uniform $\lambda$ across the inner disk and treating the LVA as a cold wind ($\beta_{\rm LVA}=0$), the $\beta$ needed to reach the observed $v_z$ of the MVA is $\beta_{\rm MVA} = v_z^2\,r_0/GM_\star - (2\lambda-3)$. Adopting $\lambda$ in the LVA range, we find $\beta_{\rm MVA} \in [0.4$-$0.9]$ for $r_0=2\,R_\star$ and $\beta_{\rm MVA} \in [0.6$-$1.4]$ for $r_0=3\,R_\star$. 
    Allowing the LVA to be warm ($\beta_{\rm LVA}>0$) would proportionally reduce the required $\beta_{\rm MVA}$, but the qualitative conclusion remains unchanged: either a larger $\lambda$ or additional pressure support is needed to justify the higher observed $v_{\rm z}$ of the MVA. 

    In summary, a consistent scenario is that the inner disk launches a warm, magneto–centrifugal disk wind with a narrow spread in opening angle ($\theta_{\rm s}\sim 12^\circ$--$14^\circ$), producing two persistent absorption components (LVA and MVA). The different observed velocities can be explained if the MVA is launched from smaller $r_0$, close to $R_{\rm T}$, and has either a higher $\lambda$ or an higher heat content at its base. The observed differences among the absorption components indicate a velocity-stratified disk wind near the star, disfavoring an X-wind scenario \citep{Shu+1994}, in which the outflow is launched from a narrow annulus at the truncation radius and therefore carries a single specific angular momentum and lever arm.

    \subsection{Spinning down RU Lup: efficiency of magnetospheric ejections}
    \label{MEs_efficiency}

    A main result of Paper\,I was the determination of the location of the disk truncation radius, $R_{\rm T}$, of RU~Lup from the analysis of the light curve obtained with the \textit{Transiting Exoplanet Survey Satellite} \citep[TESS,][]{Ricker+2014}. We derived $R_{\rm T} \approx 2 ~ R_{\star}$ and showed that $R_{\rm T}$ varies with the accretion rate, $\dot{M}_{\rm acc}$. We estimated $\dot{M}_{\rm acc} = 1.47 \times 10^{-7} ~ M_{\odot}~\rm{yr}^{-1}$ during the TESS Sector\,65 observation. Together with $M_{\star}$ (Table~\ref{tab:stellar_pars}), these measures allow us to compute the accretion torque,
    \begin{equation}
        \tau_{\rm acc} = \dot{M}_{\rm acc} \sqrt{GM_{\star} R_{\rm T}}
    \end{equation}
    \citep[e.g.,][]{MattPudritz2005a, Pantolmos+2020}. We obtain $ \tau_{\rm acc} = 9.46 \times 10^{43} ~ \rm{g~cm^2~s^{-2}}$ or, in more convenient units, $\tau_{\rm acc} = 3.2 \times 10^{-7} ~ M_{\odot}~\rm{yr^{-1}}~\rm{AU}~\rm{km~s^{-1}}$.
    
    This torque spins up the star on a characteristic timescale $t_{\rm acc} = I_{\star}\Omega_{\star}/\tau_{\rm acc}$, where $I_{\star} = (2/5)M_{\star}R_{\star}^2$ is the stellar moment of inertia and $\Omega_{\star} = 2\pi/P_{\star}$ is the stellar angular velocity.
    Using $P_{\star} = 3.71$~days (Table~\ref{tab:stellar_pars}), we derive $t_{\rm acc} \approx 3.57 \times 10^4$~yr for RU~Lup. Since the star has an age of $\sim 2-3$~Myr \citep{Herczeg+2005}, it should already rotate at break-up velocity. This implies that an efficient spin-down mechanism must be at work, removing angular momentum from the star-disk system via outflows.

    In Sect.~\ref{LVA_MVA_discussion} we showed that the MVA likely traces an ejection launched from very small $r_0$, in a region of the disk that is magnetically connected to the star. Its kinematics appears to respond to the accretion state: the toroidal velocity is higher in ES~22.3 than in ES~22.4 (Table~\ref{tab:LVA_MVA_summary}), and Paper~I reports a larger veiling fraction in ES~22.3, indicating stronger accretion. This correlation supports the idea that the outflow traced by the MVA may act as an eﬃcient channel for angular momentum removal from the star-disk system. Moreover, our analysis indicates that the winds traced by the LVA and MVA are substantially mass-loaded and warm, as expected for outflows launched from the inner disk. This picture is consistent with models in which the magnetosphere episodically ejects blobs of gas that contribute to stellar spin-down during enhanced accretion phases \citep[e.g.,][]{ZanniFerreira2013}. In this framework, our measurements of $\dot{M}_{\rm MVA}$ and $\ell$ allow us to quantitatively address the spin-down problem.

    To assess whether the outflow traced by the MVA can counteract the spin-up torque from accretion, we express the wind torque as a function of the fraction of sodium in the neutral state, $\chi$. {Combining $\dot{M}_{\rm MVA}$ with $\ell$ obtained from the ES 22.3 spectrum (Table~\ref{tab:LVA_MVA_summary}), the upper limit on the MVA torque $\tau_{\rm MVA} = \dot{M}_{\rm MVA}~\ell$ is $5.02 \times 10^{-11} \, \chi^{-1} \; M_\odot~\rm{yr}^{-1} ~ \rm{AU\,km\,s^{-1}}$.}

    Balancing $\tau_{\rm MVA}$ with the accretion torque requires a neutral fraction $\chi_{\rm b} \gtrsim 1.57 \times 10^{-4}$, that is, $\dot{M}_{\rm MVA} \lesssim 2.1 \times 10^{-8} ~ M_{\odot} ~ \rm{yr}^{-1}$, which is a fraction $\sim 20\%$ of $\dot{M}_{\rm acc}$. If the neutral fraction increases to $\chi = 3.3 \times 10^{-4}$ so that we get $\dot{M}_{\rm MVA} \approx 0.1 \dot{M}_{\rm acc}$ (Sect.~\ref{DACs_mass_loss}), the MVA wind still carries away $\lesssim 50\%$ of the accretion torque, indicating a substantial contribution to angular momentum removal.
    Together with the modest lever arm, these results reinforce a warm, mass-loaded disk wind solution capable of regulating the stellar spin.
    
    \section{Conclusions}
    \label{conclusions}
\begin{figure}
        \centering
        \includegraphics[width=\linewidth]{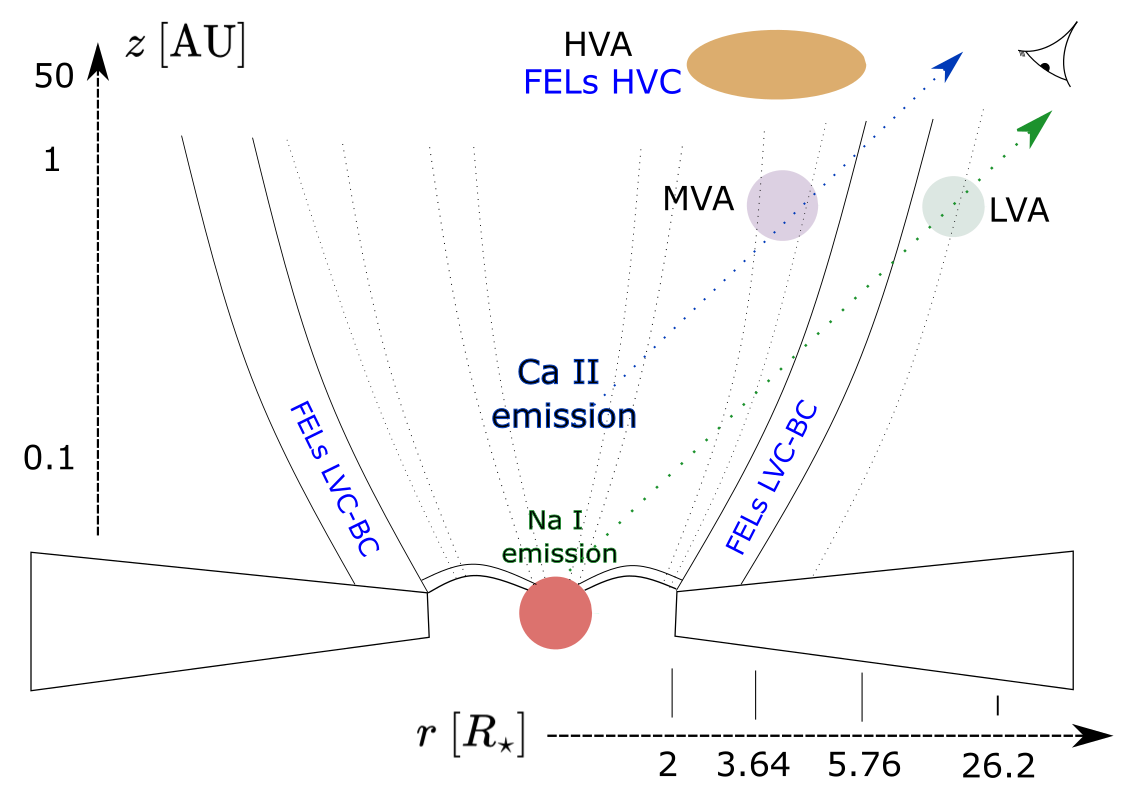}
        \caption{Sketch of the outflow structure of RU~Lup inferred from our analysis. The figure is not to scale.}
        \label{fig:outflow_final_sketch}
    \end{figure}

    In this work we used the high resolution ESPRESSO spectra of RU~Lup, obtained as part of the PENELLOPE program \citep{Manara+2021}, to study the structure of the outflow of this strongly accreting CTTS. The observations trace the outflow close to the launching region and constrain the spin-down mechanism. Specifically, we combined the information that can be extracted from the forbidden emission lines with the study of absorption components in the permitted resonance lines of the \ion{Na}{i} D$_1$\&D$_2$ and \ion{Ca}{ii} H\&K doublets.
    Figure~\ref{fig:outflow_final_sketch} sketches the multi-component outflow and the locations of the observed absorptions.
    
    The main results can be summarized as follows.
    \begin{itemize}
        \item The LVC–BC of the forbidden emission lines traces an MHD wind launched from the inner disk, which we identify with the conical wind predicted by MHD simulations \citep{Romanova+2009}.
        \item The HVA component is connected to the HVC of the forbidden lines; both originate in the low-density outer jet at distances $\gtrsim 50$~AU.
        \item In the 2022 ESPRESSO spectra of RU~Lup, the MVA and LVA sometimes display a double-dipped absorption consistent with rotation. We developed a method to disentangle the vertical ($v_{\rm z}$) and toroidal ($v_{\phi}$) velocities in these DACs and to infer the launching radius $r_0$, magnetic lever arm $\lambda$, and mass-loss rate $\dot{M}_{\rm wind}$.
        \item The LVA and MVA trace a warm, highly mass-loaded disk wind launched from the inner disk ($r_0 \lesssim 6.76\,R_{\star}$) with low magnetic lever arms \citep{CasseFerreira2000B}. The two components mainly differ in their observed poloidal velocities ($v_{\rm z}$ is about twice as large in the MVA as in the LVA) and in the weakness of the LVA in \ion{Ca}{ii}.
        \item The LVA is explained as an outer absorbing shell with $\lambda_{\rm LVA}\!\in\![1.56,\,1.93]$ and a low covering of the \ion{Ca}{ii} line-emitting region. We attribute its weakness in \ion{Ca}{ii} to geometry (small line-emitter covering fraction) rather than to a lower ionization than the MVA.
        \item The MVA traces a distinct inner wind layer forming near the truncation radius $R_{\rm T}$, whose higher $v_{\rm z}$ is compatible with either a slightly larger lever arm, $\lambda_{\rm MVA}\!\in\![1.75,\,3.20]$, or with additional thermal support at the base, parameterized by $\beta\!\in\![0.4,\,0.9]$.
        \item The LVA/MVA contrast arises from different covering fractions and stratification in a warm disk wind launched across a narrow annulus near $R_{\rm T}$, disfavoring a pure X-wind scenario \citep[e.g.,][]{Shu+1994}.
        \item We evaluated the MVA wind torque as a function of the sodium neutral fraction $\chi$, and showed that the MVA removes a substantial fraction of the accretion spin-up torque for plausible ionization levels in the inner disk.
    \end{itemize}

    Future studies should aim at extending this analysis to other systems. While the low inclination of RU~Lup provides a direct view of the outflow, systems observed at higher inclinations are expected to exhibit a higher radial velocity gradient due to the toroidal motion, with the advantage of more effectively separating the double-dipped structure of the absorption components. Additionally, time-resolved spectroscopic observations across different timescales would enable a detailed investigation of the variability of the different components of the outflow and their connection to the accretion process.
    
    \bigskip

    \begin{acknowledgements} 
        The authors thank the anonymous referee for their critical review of this manuscript.
        This work has been supported by Deutsche Forschungsgemeinschaft (DFG) in the framework of the YTTHACA Project (469334657) under the project codes STE 1068/9-1 and MA 8447/1-1.
        AF acknowledges financial support from the Large Grant INAF 2022 "YSOs Outflows, Disks and Accretion: towards a global framework for the evolution of planet forming systems" (YODA).
        CFM and JCW are funded by the European Union (ERC, WANDA, 101039452). Views and opinions expressed are however those of the author(s) only and do not necessarily reflect those of the European Union or the European Research Council Executive Agency. Neither the European Union nor the granting authority can be held responsible for them. 
        JFG was supported by Fundação para a Ciência e Tecnologia (FCT) through the research grants UIDB/04434/2020 and UIDP/04434/2020.
        The authors acknowledge the use of the electronic bibliography maintained by the NASA/ADS\footnote{\url{https://ui.adsabs.harvard.edu}} system.
    \end{acknowledgements}

    \bigskip

    \bibliographystyle{aa} 
    \bibliography{RU_LUP.bib} 
	    
    \begin{appendix}
        \section{Log of spectroscopic observations}
        \begin{table}
            \centering
        	\caption{Log of the spectroscopic observations.} 
        	\begin{tabular}{ccccccc}
                \hline
                ID & MJD ($- 52300$ days) & S/N & $t_{\rm exp}$~(s) \\
                \hline 
                ES 21.1 & 7149.00 & 55 & 600 \\ 
                ES 21.2 & 7158.06 & 32 & 400 \\
                ES 22.1 & 7501.04 & 47 & 600 \\
                ES 22.2 & 7502.06 & 46 & 600 \\
                ES 22.3 & 7504.10 & 50 & 600 \\
                ES 22.4 & 7507.10 & 20 & 600 \\
                ES 22.5 & 7514.04 & 46 & 600 \\
                \hline
        	\end{tabular}
        	\tablefoot{The spectra are labelled as in Paper I. The resolving power of the ESPRESSO spectra is $R = 140000$.}
        	\label{tab:log_specobs} 
        \end{table}
        
        Table~\ref{tab:log_specobs} reports the spectroscopic observations used in this work.

        \section{Photospheric subtraction in the forbidden emission lines}
        \label{photospheric_subtraction}

        \begin{figure*}
            \centering
            \includegraphics[width=\linewidth]{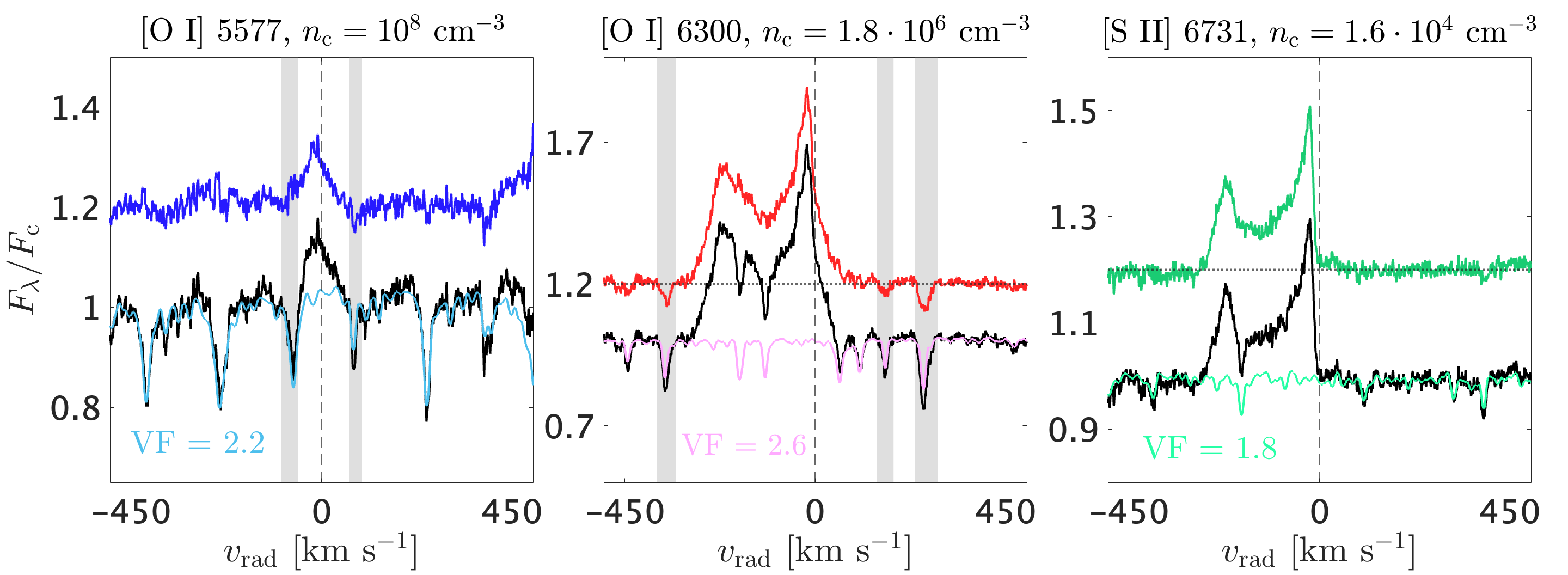}
            \caption{Photospheric subtraction procedure for the [\ion{O}{i}] 5577, [\ion{O}{i}] 6300, and [\ion{S}{ii}] 6731 lines in the ES~22.5 spectrum of RU~Lup. The black lines are the observed spectra. The colored lines superposed on the observed spectra are from the K7 template for the photospheric spectrum of RU~Lup, veiled to match the depth of the photospheric lines. Above these spectra are the photospheric-subtracted spectra, shifted vertically by $0.2$ for clarity. The shaded areas mark the region where the photospheric spectrum is not completely removed.}
            \label{fig:FELs}
        \end{figure*}
         
        The line profiles of the forbidden emission lines of RU~Lup are severely blended with the absorption lines from the photosphere of the star. To study the kinematics of the outflow, it is important to remove the stellar contribution from the profiles of the {forbidden lines}. To this end, we used the template spectrum that we have identified in Paper\,I as best fitting the photospheric spectrum of RU\,Lup.  
        The properties of the template were obtained using the ROTFIT code \citep{Frasca+2015}, which uses a library of \textit{High Accuracy Radial velocity Planet Searcher} \citep[HARPS,][]{Mayor+2003} spectra from the ESO Archive to fit the photospheric spectrum. The details of the procedure to obtain the template are reported in Paper\,I. 
        
        For each emission line, we adjusted the veiling fraction (VF) in its vicinity to match the depth of the photospheric lines. The procedure and the result of the photospheric subtraction are shown in Fig.~\ref{fig:FELs}. Due to the effect of line filling emission in the photospheric lines, which we analyzed in Paper I, the photospheric subtraction has some imperfections. The absorption lines that sit on top of the [\ion{O}{i}] 6300 and [\ion{S}{ii}] 6731 lines have been correctly removed, but the wings of the [\ion{O}{i}] 5577 line are still slightly contaminated, especially the red one. 

        \section{Stellar and accretion parameters}
        For the convenience of the reader, we report in Table~\ref{tab:stellar_pars} the stellar parameters of RU~Lup which are needed for the analysis of the outflow, together with the position of the corotation radius ($R_{\rm co}$) and the truncation radius ($R_{\rm T}$). 

        \begin{table}
        	\centering
        	\caption{Stellar and accretion parameters of RU\,Lup.} 
        	\begin{tabular}{ccc}
            	\hline
            	Parameter & Value & Ref. \\
            	\hline
                    $d$ & $158.9 \pm 0.7$~pc & [1] \\
                    SpT & K7 & [2] \\
                    $T_{\rm eff}$ & $4250 \pm 60$~K & [3] \\
                    $v \sin i$ & $8.6 \pm 1.4~\rm{km~s^{-1}}$ & [3]  \\
                    $i_{\star}$ & $16 \pm 5~{}^{\rm o}$ & [3] \\ 
                    $R_{\star}$ & $2.27 \pm 0.52$~$R_{\odot}$ & [3] \\
                    $M_{\star}$ & $0.55 \pm 0.13$~$M_{\odot}$ & [4] \\
                    $L_{\star}$ & $1.46 \pm 0.67$~$L_{\odot}$ & [4] \\
                    $P_{\star}$ & $3.71 \pm 0.01$ d & [5] \\  
                    $R_{\rm co}$ & $\sim 3.64 ~ R_{\star}$ & [3] \\
                    $R_{\rm T}$ & $\sim 2 ~ R_{\star}$ & [3] \\
        	    \hline
        	\end{tabular}
        	\tablefoot{References: [1] \citet{GaiaDR3}; [2] \citet{Alcala+2017}; [3] Paper\,I; [4] \citet{Manara+2023}; [5] \citet{Stempels+2007}.} 
        	\label{tab:stellar_pars} 
        \end{table}
      
        \section{Column density of the absorption components}
        \label{DACs_EW}
        The LVA is detected only in the \ion{Na}{i} lines. This suggests that most of the gas in the absorbing region is cold, hence neutral and in the ground state. Since \ion{Na}{i}~D$_2$ is a resonance line, we can estimate the column density, $N$, of the absorbing material without any assumption on the temperature. 

\begin{figure}
            \centering
            \includegraphics[width=\linewidth]{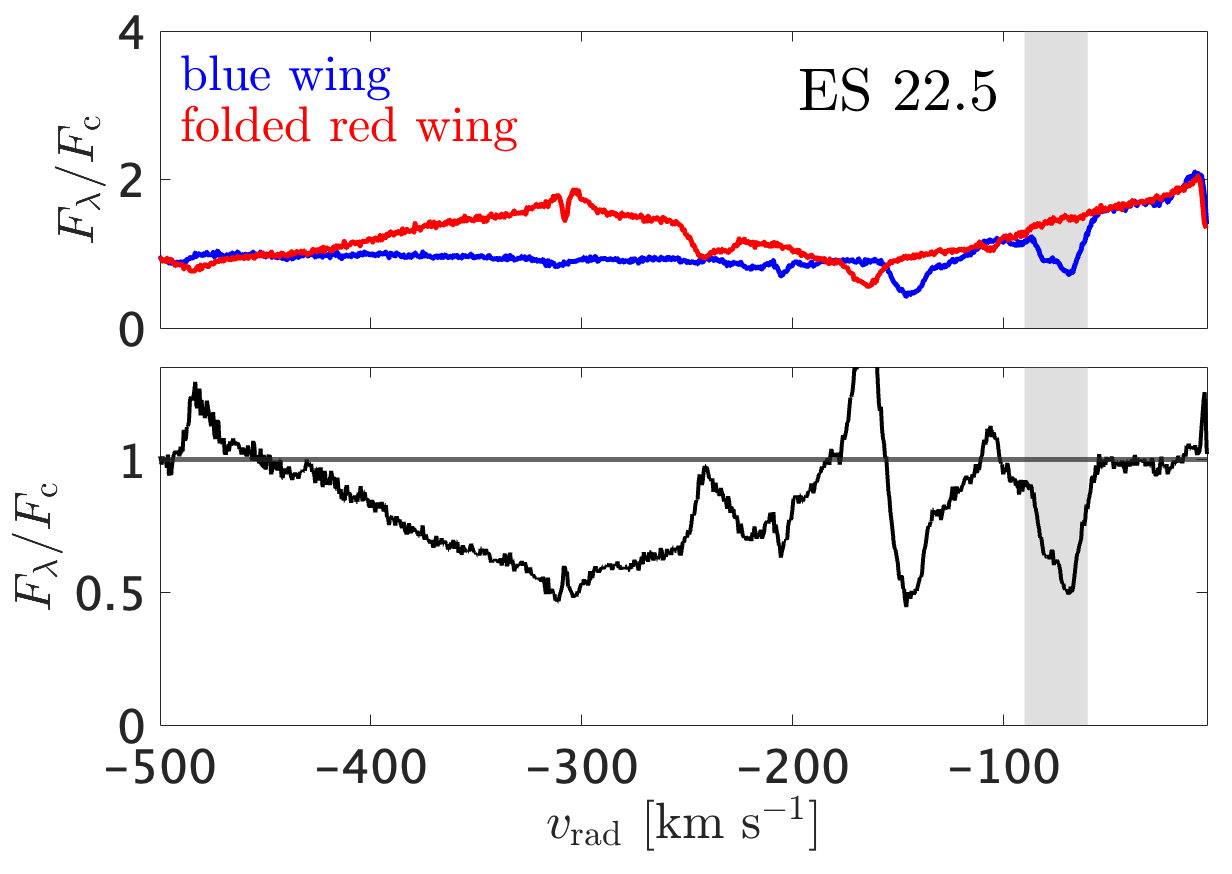}
            \caption{Extraction of the LVA component of the \ion{Na}{i}~D$_2$ line in the ES~22.5 spectrum. The top panel shows the red wing line folded onto the blue wing. The bottom panel shows the ratio between the blue wing and the red wing, from which we computed the optical depth and EW. The shaded area marks the velocity extension of the LVA, as defined in Fig.~\ref{fig:CaII_vs_NaI}.}
            \label{fig:LVA_EW}
        \end{figure}

        The easiest way to do this is computing the equivalent width (EW) of the line. In the optically thin limit, the EW is linearly related to the column density of sodium atoms. To this end, we need a model for the un-absorbed emission profile. We used the red wing of the \ion{Na}{i}~D$_2$ line as a model for the blue wing emission, by folding the D$_2$ line profile around the line center as shown in the top panel of Fig.~\ref{fig:LVA_EW}. {This was possible because in the region of the LVA the red portion of the line is free of absorption lines from the stellar photosphere.}
   
        The normalized profile $F_{\lambda}/F_{\rm c}$ (bottom panel) is derived by dividing the blue wing by the folded red wing. 
        The optical depth, assuming complete covering of the emitting region, is $\tau_{\lambda} = - \ln{(F_{\lambda}/F_{\rm c})}$. At line minimum we obtain $\tau_{\lambda} = 0.71$, confirming that the gas is not optically thick. We derived EW by integrating the normalized profile in the $v_{\rm rad}$ range of the LVA, that is, $-90 ~ \rm{km~s^{-1}} \leq v_{\rm rad} \leq -60 ~ \rm{km~s^{-1}}$. The result is $\rm{EW} = 0.19$~{\AA}.
        We applied this procedure to the ES~22.5 spectrum. For the other spectra, the red wing did not accurately reproduce the unabsorbed LVA profile.

        We converted the EW to $N$ by inverting the formula
        \begin{equation}
            \frac{\rm EW}{\lambda_0} = (8.85 \times 10^{-21}) \, N \lambda_0 f,
        \end{equation}
        where $\lambda_0$ and $f$ are the rest wavelength in {\AA} and the oscillator strength of the transition, respectively, and $N$ is in $\rm{cm}^{-2}$ \citep[e.g.,][]{Spitzer+1998}. For the D$_2$ line we used $\lambda_0 = 5889.95$~{\AA} and $f = 0.641$ from the NIST Atomic Spectra Database\footnote{\url{https://physics.nist.gov/PhysRefData/ASD/lines_form.html}}. 
        
        The result is a column density of atoms in the ground state of the \ion{Na}{i} ion, $N_0$, of $9.95 \times 10^{11} ~ \rm{cm^{-2}}$. 
        Assuming that all sodium is neutral and in the ground state, then $N_0$ represents the total column density of sodium, $N_{\rm Na}$. 
        Using the solar abundances from \citet{Asplund+2009}, i.e., $N_{\rm Na} / N_{\rm H} = 1.74 \times 10^{-6}$, we obtain the column density of hydrogen, $N_{\rm H} = 5.49 \times 10^{17} ~ \rm{cm^{-2}}$.

        We repeated this procedure for the MVA of the \ion{Na}{i}~D$_1$ line in the ES~22.4 spectrum. We obtained $\rm{EW} = 0.20$~{\AA}, which gives a column density of the atoms in the ground state of the \ion{Na}{i} ion of $N_0 = 2.07 \times 10^{12} ~ \rm{cm^{-2}}$. To convert this value into a total hydrogen column density, we must account for the excitation and ionization state of sodium in the MVA gas, which is expected to be partially ionized. The total hydrogen column density can be expressed as:
        \begin{equation}
            N_{\rm H} = N_0 \cdot \left( \frac{N_{\rm H}}{N_{\rm Na}} \right) \cdot \left( \frac{N_{\rm Na}}{N_{\rm Na\,I}} \right) \cdot \left( \frac{N_{\rm Na\,I}}{N_0} \right).
        \end{equation}
        The first term, $N_{\rm H} / N_{\rm Na} = 5.75 \times 10^5$, is the inverse of the sodium abundance. The second term, $N_{\rm Na} / N_{\rm Na\,I}$, describes the ionization state of the gas and is denoted as $\chi^{-1}$, where $\chi \equiv N_{\rm Na\,I} / N_{\rm Na}$. This ratio can be estimated from the Saha equation for a given temperature and electron density. The third term, $N_{\rm Na\,I} / N_0$, accounts for the population of sodium atoms in excited levels and is given by the Boltzmann distribution, $N_{\rm Na\,I}/N_0 = U(T)/g_0$.
        Here $g_0 = 2$ is the statistical weight of the ground state, and $U(T)$ is the partition function of \ion{Na}{i} at temperature $T$. The partition function is only weakly dependent on temperature, with $U(T) \approx 2-2.5$ for $T$ between $4000$ and $8000$~K. This yields a correction factor $N_{\rm Na\,I} / N_0 \approx 1.0-1.25$. The most uncertain parameter in this conversion is the {fraction of sodium atoms in the neutral state} $\chi$. 
        We write the hydrogen column density associated with the MVA as
        \begin{equation}
            N_{\rm H} \approx N_0 ~ \frac{5.75 \times 10^5}{\chi} = \frac{1.16 \times 10^{18} ~ \rm{cm^{-2}}}{\chi}.
        \end{equation}
        
        \section{Synthetic absorption profiles from a rotating and outflowing ring}
        \label{ring_model}

        \begin{figure}
            \centering
            \includegraphics[width=0.75\linewidth]{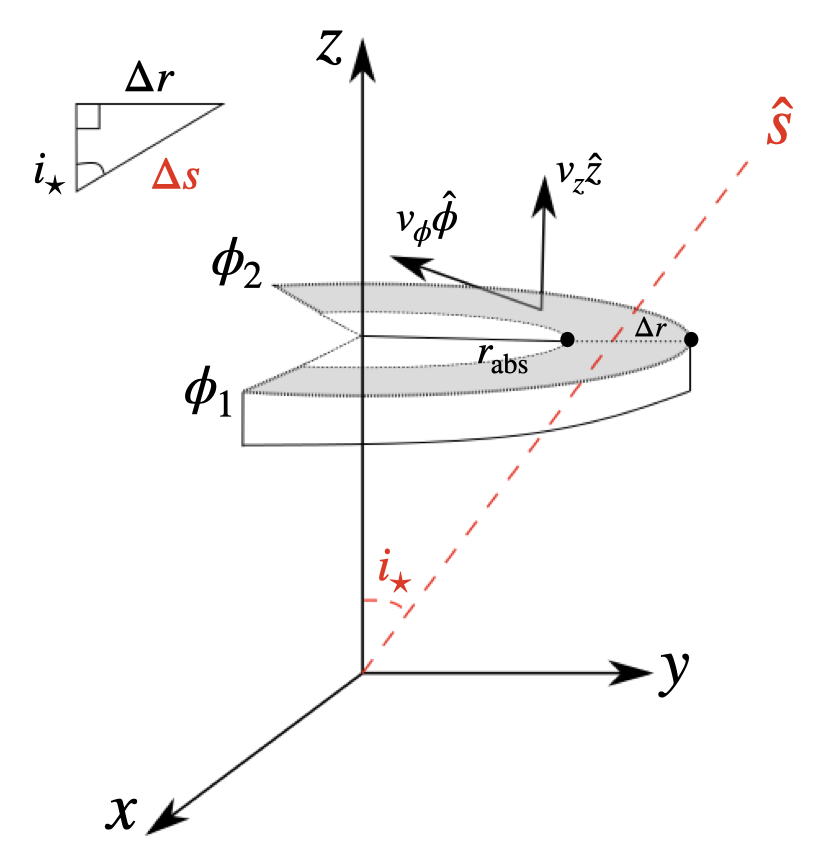}
            \caption{Sketch of {the geometry of the rotating and outflowing sector of a ring in cylindrical coordinates.}}
            \label{fig:ring_abs_cylindrical}
        \end{figure}

        {In this Appendix we highlight the details of the model used to interpret the {discrete absorption components} in terms of a rotating and outflowing structure.}
        We considered a cylindrical coordinate system $(r, \phi, z)$, where $r$ represents the distance from the rotation axis in the $x$-$y$ plane, $\phi$ is the azimuthal angle measured relative to the x-axis, and $z$ corresponds to the rotation axis (Fig.~\ref{fig:ring_abs_cylindrical}). In cylindrical coordinates the velocity vector is $\vec{v} = v_r \hat{\vec{r}} + v_{\phi}\hat{\vec{\phi}} + v_z \hat{\vec{z}}$. The unit vectors are given by $\hat{\vec{r}} = (\cos \phi, \sin \phi, 0)$,  $\hat{\vec{\phi}} = (-\sin \phi, \cos \phi, 0)$, and $\hat{\vec{z}} = (0, 0, 1)$ in Cartesian coordinates. The radial velocity is $v_{\rm rad} = - \vec{v} \cdot \hat{\vec{s}}$, where $\hat{\vec{s}} = (\sin i_{\star}, 0, \cos i_{\star})$ represents the observer's line of sight. The dot product yields
        \begin{equation}
            v_{\rm rad} = - v_r \cos \phi \sin i_{\star} + v_{\phi} \sin \phi \sin i_{\star} - v_z \cos i_{\star}.
            \label{vrad_toroidal_wind}
        \end{equation}
        Assuming that the poloidal flow is predominantly parallel to the z-axis, we can neglect $v_r$ with respect to $v_z$.
        Then, the shape of the line profile depends on the angles $\phi_1$ and $\phi_2$ which limit the sampled region in azimuth. 
        
        To compute the absorption profiles produced by the ring, we generated an array of azimuth angles ($\phi$) between $\phi_1$ and $\phi_2$, and calculated the local optical depth as
        \begin{equation}
            d\tau_v(\phi) = \tau_0 ~ \frac{1}{\sqrt{2 \pi}\sigma} \exp{\left[- \frac{(v - v_{\rm rad}(\phi))^2}{2 \sigma^2} \right]}.
        \end{equation}
        where $v_{\rm rad}(\phi)$ and $\sigma$ represent the center and standard deviation of the Gaussian function, and $\tau_0$ is a scaling parameter. 
        Using the radial velocity law of Eq.~\eqref{vrad_toroidal_wind}, we summed all contributions in $\phi$ to produce the total optical depth $\tau_v$ as a function of velocity.
        Then, we computed the resulting normalized line profile as
        \begin{equation}
            F_v = 1 - \rm{CF}[1 - \exp(-\tau_v)],
        \end{equation}
        where CF is the covering factor, that is, the fraction of the emission region that is covered.
        
\begin{figure}
            \centering
            \includegraphics[width=\linewidth]{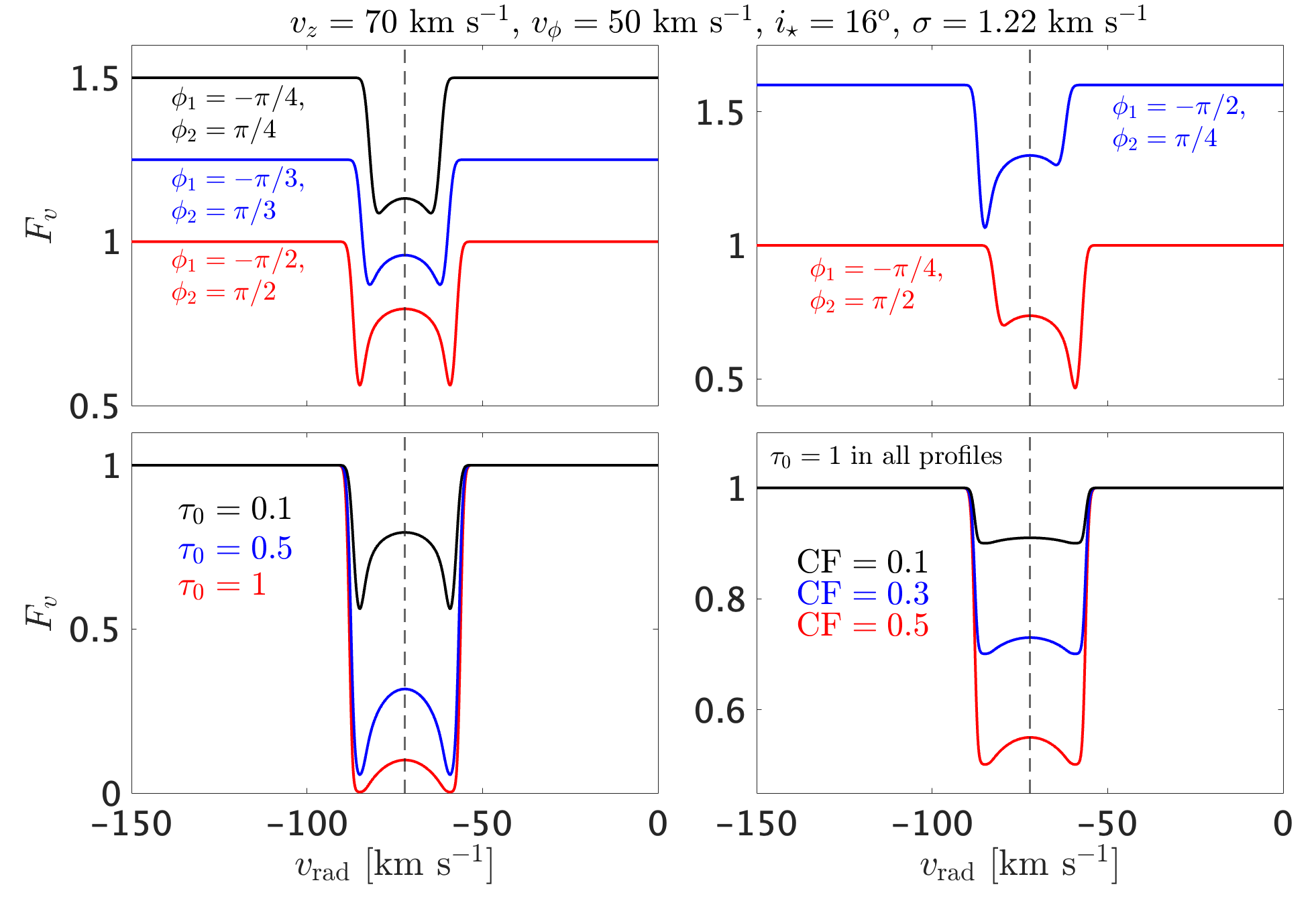}
            \caption{Gallery of absorption profiles for a vertically-narrow region of a wind which is extended in azimuth. In all plots, $v_z = 70 ~\rm{km~s^{-1}}$, $v_{\phi} = 50 ~\rm{km~s^{-1}}$, $i_{\star} = 16^{\rm o}$, and $\sigma = 1.22 ~ \rm{km~s^{-1}}$. 
            If not reported, the parameters are $\phi_1 = -\pi/2$, $\phi_2 = \pi/2$, $\tau_0 = 0.1$, and $\rm{CF} = 1$. 
            The vertical dashed lines is are $-v_z \cos i_{\star}$.}
            \label{fig:jet_abs_profiles}
        \end{figure}

        The shape of the line profile depends on the velocities $v_z$ and $v_{\phi}$, the angles $\phi_1$ and $\phi_2$, and the scaling parameters $\sigma$, $\tau_0$, and CF.
        Figure~\ref{fig:jet_abs_profiles} displays examples of absorption profiles produced with this model for typical velocities observed in the LVA, that is, $v_z = 70 ~\rm{km~s^{-1}}$ and $v_{\phi} = 50 ~\rm{km~s^{-1}}$. We assumed that the local broadening is thermal, that is, $\sigma = (2k_B T/Am_{\rm H})^{1/2}$, where $k_{\rm B}$ is the Boltzmann constant, $T$ is the temperature, $m_{\rm H}$ is the hydrogen mass, and $A$ is the atomic number. In Fig.~\ref{fig:jet_abs_profiles}, we fixed $T = 1000$~K and $A = 11$ (sodium), which yielded $\sigma = 1.22 ~ \rm{km~s^{-1}}$.

        The profiles have two main characteristics. First, they are double-dipped and the velocities of the dips are $v_1 = -v_{z} \cos i_{\star} + v_{\phi} \sin \phi_1 \sin i_{\star}$ and $v_2 = -v_{z} \cos i_{\star} + v_{\phi} \sin \phi_2 \sin i_{\star}$. Second, when the profiles are symmetric, the center is at $v_0 = -v_{z} \cos i_{\star}$.

        \section{Inferring the outflow properties from the ring fit}    
        {Our fit of the double-dipped absorption components (Sect.~\ref{poloidal_toroidal_decomposition}) disentangles $v_\phi$ from $v_{\rm z}$. From these values, we developed a procedure to infer the launching radius $r_0$ and the magnetic lever arm $\lambda$ of the outflow under the assumption of a steady, axisymmetric, cold magneto-centrifugal wind \citep{Ferreira+2006}.}

        \subsection{Geometrical estimate of the absorbing radius}
        \label{rabs_derivation}
\begin{figure}
            \centering
            \includegraphics[width=\linewidth]{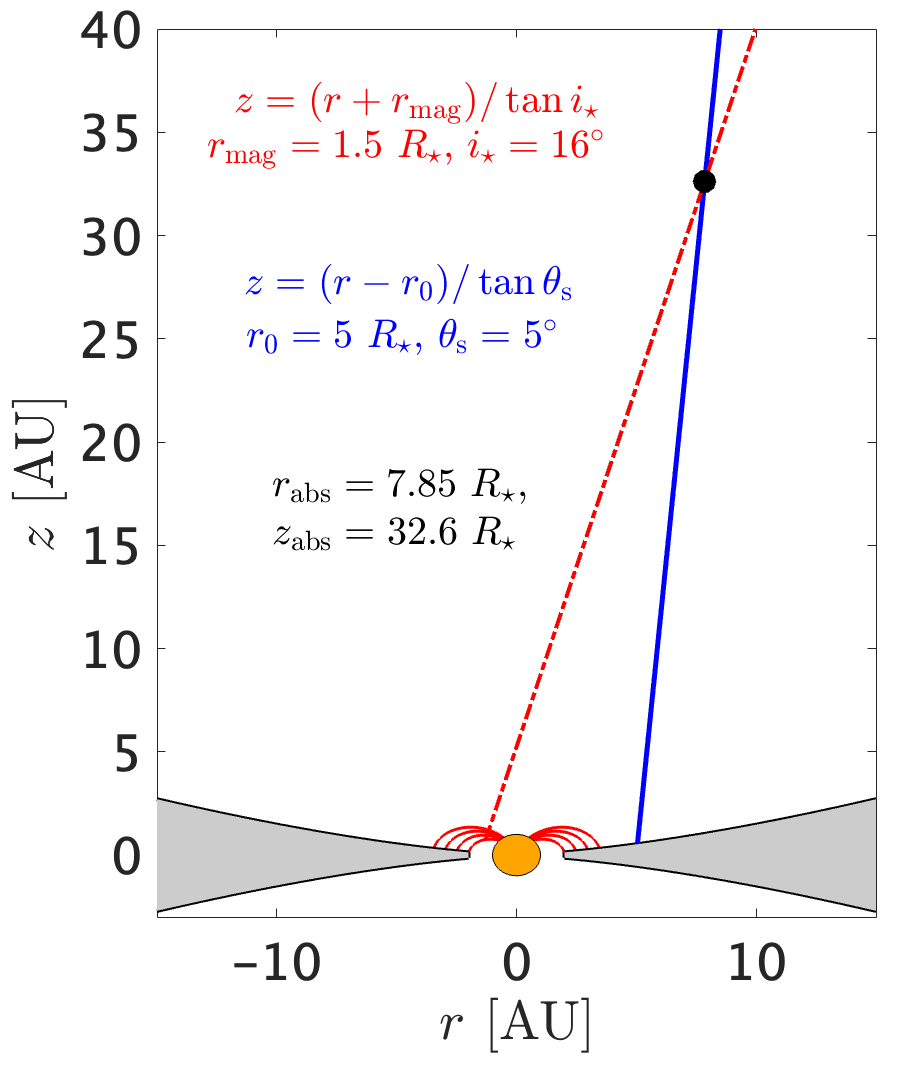}
            \caption{{Example of the geometrical model used to estimate the absorbing radius of the {discrete absorption components}. The red dashed line represents the line of sight to the magnetosphere. The blue line shows a streamline launched from the disk. The black dot marks the intersection point between the two lines, which defines the location of the absorbing gas.}}
            \label{fig:abs_radius_geometry}
        \end{figure}

        {Although we do not directly know $r_{\rm abs}$, the absorbing parcel of gas must lie at the intersection between a streamline and the line of sight (LOS) to the line-emitting region. The streamline can be parameterized as a straight line that originates from the disk at cylindrical radius $r_0$ and has an inclination $\theta_s$ (measured from the disk normal), i.e., $z = (r - r_0)/\tan \theta_{\rm s}$. The line emission is assumed to come from the blueshifted portion of the magnetosphere, located at cylindrical radii $r_{\rm mag} \in [0, R_{\rm T}]$, where $R_{\rm T}$ is the disk truncation radius. Therefore, the LOS has the expression $z = (r + r_{\rm mag})/\tan i_{\star}$.
        By solving for the intersection between these two lines, we find
        \begin{equation}
        r_{\rm abs} = \frac{ r_0 \tan i_\star + r_{\rm mag}\,\tan\theta_s }{ \tan i_\star - \tan\theta_s }
        \label{eq:rabs_full}
        \end{equation}
        The only physical solutions are the ones in which absorption takes place above the disk plane ($z > 0$). This condition implies that the streamline must be less inclined than the line of sight, that is, $\theta_s < i_{\star}$. A schematic view of the adopted geometry is shown in Fig.~\ref{fig:abs_radius_geometry}.}

        {For the following, we define the geometry factors
        \begin{equation}
        K \;\equiv\; \frac{\tan i_\star}{\tan i_\star - \tan\theta_s},
        \qquad
        \alpha \;\equiv\; \frac{\tan\theta_s}{\tan i_\star - \tan\theta_s},
        \label{eq:Kalpha}
        \end{equation}
        so that $r_{\rm abs} = K\,r_0 + \alpha\,r_{\rm mag}$.}
        
        \subsection{Solving for the launching radius}
        \label{r0_lambda_derivation}
        {Combining the geometrical constraint of $r_{\rm abs}$ with the cold disk wind relations gives three equations for the three unknowns $r_0$, $\lambda$, and $r_{\rm abs}$, that is,
        \begin{equation}
            \begin{cases}
                2\lambda \;=\; 3 + v_z^2 r_0 / G M_\star, \\
                r_{\rm abs} = K\,r_0 + \alpha\,r_{\rm mag}, \\
                r_{\rm abs} v_{\phi} = \lambda \sqrt{GM_{\star}r_0}.
            \end{cases}
            \label{eq:cold_disk_system}
        \end{equation}
        Substituting $\lambda$ and $r_{\rm abs}$ (first and second equation of the system, respectively) into the third equation, and dividing by $\sqrt{G M_\star}$, we obtain Eq.~\eqref{eq:cubic} for $y = \sqrt{r_0}$. The positive real roots of this equation give $r_0=y^2$.} 

        \subsection{Launching properties of the outflow}
        \label{r0_vs_rmag}
\begin{figure}
            \centering
            \includegraphics[width=\linewidth]{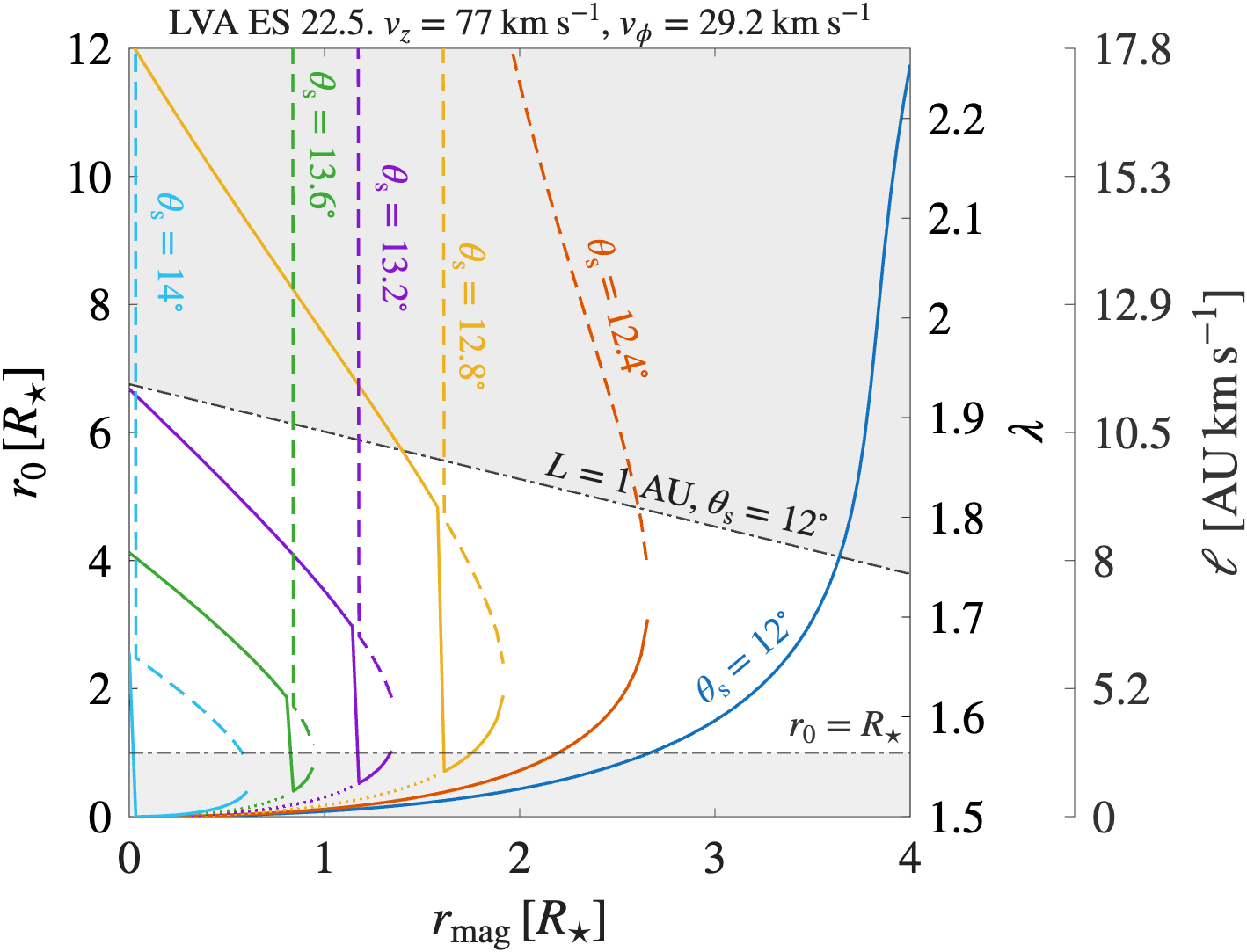}
            \caption{{$r_0$ vs. $r_{\rm mag}$ plot for the LVA in ES~22.5. Colors denote different $\theta_{\rm s}$. Solid, dashed, and dotted curves indicate distinct solution branches of Eq.~\eqref{eq:cubic}. The black dash–dotted lines mark the lower (stellar radius $R_{\star}$) and upper (given by the maximum distance at which absorption can form) limits on $r_0$. The shaded areas indicate the excluded regions. The axes on the right show the corresponding magnetic lever arm $\lambda$ and specific angular momentum $\ell$, computed using $v_{\rm z}$, $r_0$ and Eqs.~\eqref{eq:ell_cold_disk_theory} and \eqref{eq:vz_cold_disk_theory}.}}
            \label{fig:LVA_ES225_r0_vs_rmag}
        \end{figure}

\begin{figure}
            \centering
            \includegraphics[width=\linewidth]{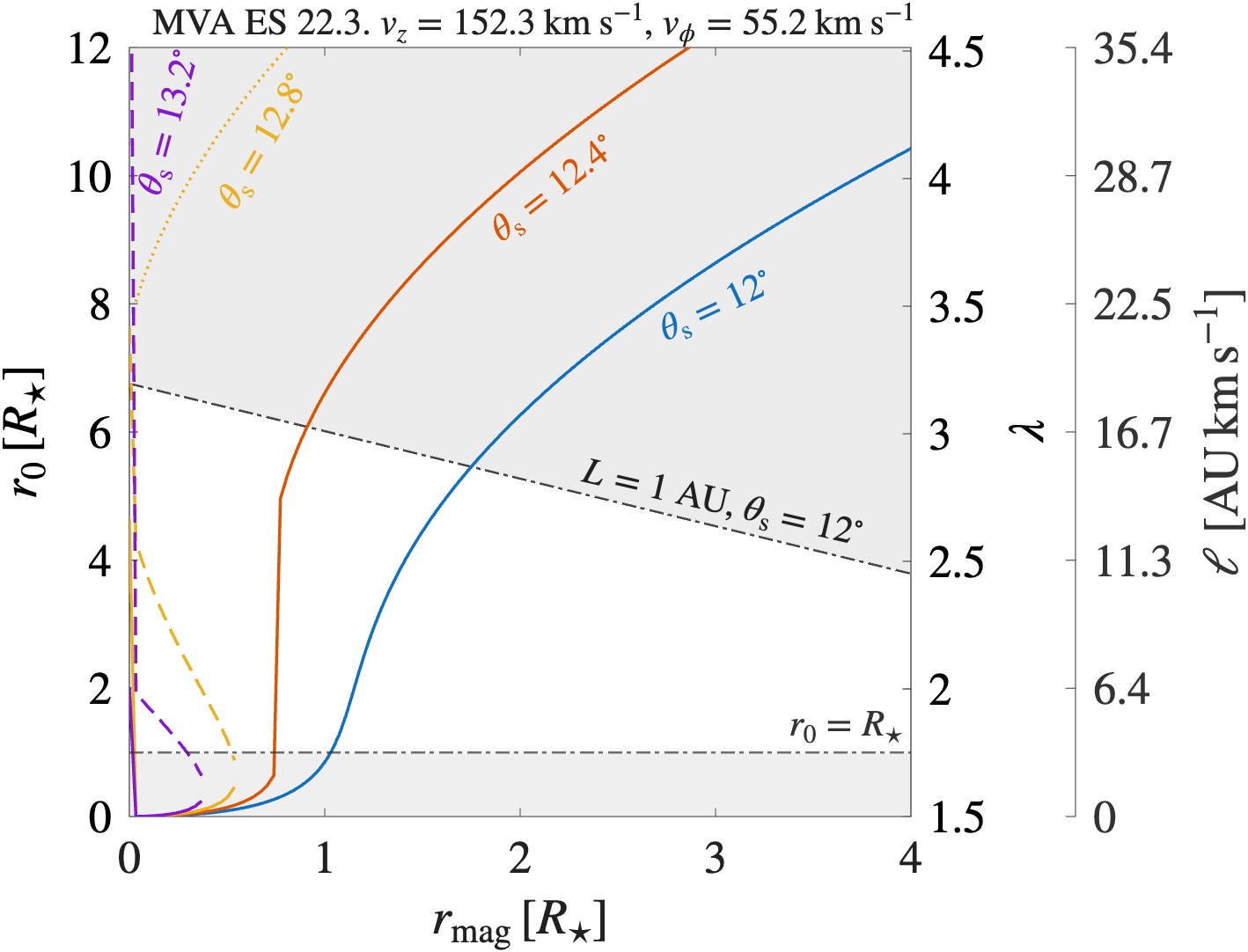}
            \caption{{Same as Fig.~\ref{fig:LVA_ES225_r0_vs_rmag}, but for the MVA in ES~22.3.}}
            \label{fig:MVA_ES223_r0_vs_rmag}
        \end{figure}

\begin{figure}
            \centering
            \includegraphics[width=\linewidth]{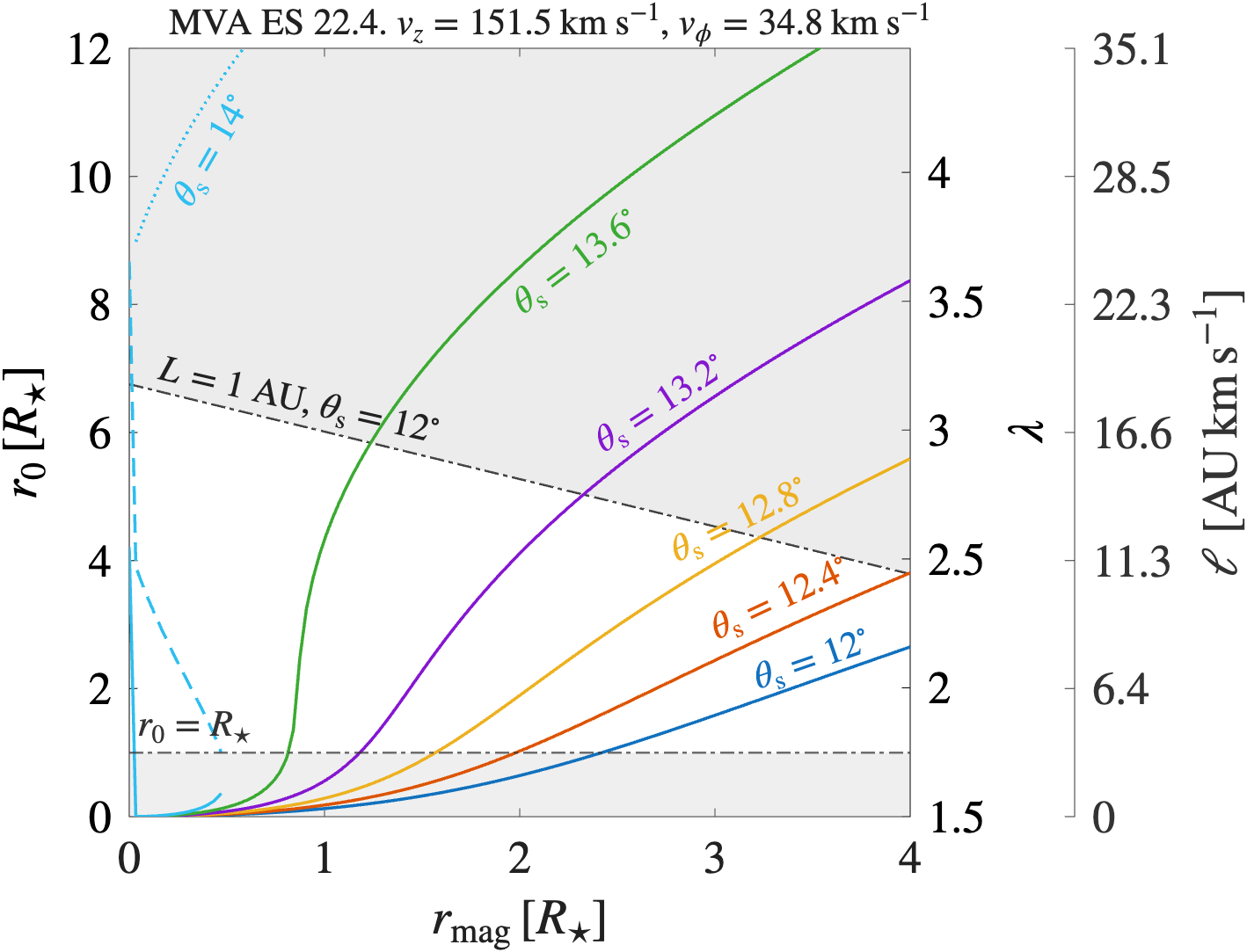}
            \caption{{Same as Fig.~\ref{fig:LVA_ES225_r0_vs_rmag}, but for the MVA in ES~22.4.}}
            \label{fig:MVA_ES224_r0_vs_rmag}
        \end{figure}

        {To explore the properties of the winds traced by the LVA and MVA, we solved Eq.~\eqref{eq:cubic} and plotted the resulting curves of $r_0$ as a function of $r_{\rm mag}$ for a grid of $\theta_{\rm s}$ values at fixed $i_{\star}=16^{\circ}$. Figure~\ref{fig:LVA_ES225_r0_vs_rmag} shows the LVA in ES~22.5, while Figs.~\ref{fig:MVA_ES223_r0_vs_rmag} and \ref{fig:MVA_ES224_r0_vs_rmag} display the MVA in ES~22.3 and ES~22.4, respectively.}

        {The parameter space is restricted by physically motivated bounds on $r_0$. We impose a lower bound at the stellar surface, $r_0>R_{\star}$, and an upper bound set by the maximum distance $L$ at which the absorption components can form. In cylindrical geometry $r_{\rm abs}=L\sin i_{\star}$, and Eq.~\eqref{eq:rabs_full} then yields an upper limit on $r_0$ as a function of $r_{\rm mag}$. Based on the results in Sect.~\ref{DACs_vs_FELs}, we adopt a conservative value of $L=1$~AU, that is, $r_{\rm abs} \lesssim 0.276 ~ \rm{AU} \simeq 26.2 ~ R_{\star}$.}
        
        {Several trends emerge from the figures. As $\theta_{\rm s}\!\to\! i_{\star}$, the solutions diverge to non-physical $r_0$. For fixed $\theta_{\rm s}$ they also rise rapidly with increasing $r_{\rm mag}$. The strongest constraint comes from the limit on $r_{\rm abs}$: for both components we obtain $r_0\!\lesssim\!6.76\,R_{\star}$. Although $\theta_{\rm s}$ is formally limited only by $\theta_{\rm s}<i_{\star}$, in practice it is confined to a narrow interval. For $r_{\rm mag}\!\lesssim\!2\,R_{\star}$, one requires $\theta_{\rm s}\!\gtrsim\!12^{\circ}$ to satisfy the lower bound on $r_0$, and $\theta_{\rm s}\!\lesssim\!14^{\circ}$ to remain below the upper bound.}
        
        Despite similar $v_{\rm z}$, the differing $v_\phi$ for the MVA in ES~22.3 and ES~22.4 shift the curves with same $\theta_{\rm s}$ into different regions of the parameter space. For example, while for the MVA in ES~22.3 the solution with $\theta_{\rm s} = 12.4^{\circ}$ has $r_{\rm mag} \lesssim 1 ~ R_{\star}$, the same solution has $r_{\rm mag} \gtrsim 2 ~ R_{\star}$ for the MVA in ES~22.4. Assuming that the absorption is formed against the magnetospheric line emission and the stellar continuum (i.e., $r_{\rm mag} \lesssim 2 R_{\star}$), this implies a increase of $\Delta \theta_{\rm s} \sim 1$-$1.5^{\circ}$ in the inclination of the streamlines between ES~22.3 and ES~22.4.
        
        Given its lower $v_{\rm z}$, the LVA requires a smaller magnetic lever arm than the MVA at a given $r_0$. We find $\lambda_{\rm LVA}\in[1.56,\,1.93]$ and $\lambda_{\rm MVA}\in[1.75,\,3.20]$.

        \section{Partial coverage of continuum and line emission}
        \label{MVA_LVA_coverage}
        
        In this Appendix we show how partial coverage of the \ion{Ca}{ii} line--emitting region can hide absorption at LVA velocities while allowing the MVA to appear strong and remain above the continuum.
        We define:
        \begin{itemize}
          \item $C$ as the \ion{Ca}{ii} continuum flux,
          \item $L(v)$ as the intrinsic \ion{Ca}{ii} line emission profile,
          \item $\tau$ as the optical depth of the absorber,
          \item $\rm{CF}_c$ as the covering fraction of the continuum source,
          \item $\rm{CF}_\ell$ as the covering fraction of the line-emitting region,
          \item $F(v)$ as the observed flux including absorption.
        \end{itemize}
        With independent partial covering of continuum and line emission, the full profile is
        \begin{multline}
            F(v) = C\!\left[(1-\mathrm{CF}_c)+\mathrm{CF}_c\,e^{-\tau}\right] \\
        + L(v)\!\left[(1-\mathrm{CF}_\ell)+\mathrm{CF}_\ell\,e^{-\tau}\right].
        \end{multline}
        The underlying emission profile at velocity $v$ is $C + L(v)$. The fractional absorption relative to this level is
        \begin{equation}
            \frac{C+L(v) - F(v)}{C + L(v)}
            =
            \big(1-e^{-\tau}\big)
            \frac{{\rm CF}_c + \frac{L(v)}{C}\rm{CF}_\ell}{1 + \frac{L(v)}{C}}.
            \label{eq:frac_abs}
        \end{equation}
        The ES~22.5 spectrum has $\mathrm{S/N}=46$, and we adopt a $3\sigma$ visibility threshold, meaning that absorption is undetected if the fraction of Eq.~\eqref{eq:frac_abs} is smaller than $\epsilon = \frac{3}{\mathrm{S/N}} \approx 0.07$. This way, we get a threshold for the overlap of the absorbing region with the line emitter, that is,
        \begin{equation}
            {\rm CF}_\ell < \frac{\epsilon\,[1+L(v)/C] - \rm{CF}_c\,(1-e^{-\tau})}{(L(v)/C)\,(1-e^{-\tau})}\,.
            \label{eq:CF_line_max}
        \end{equation}
        
        Figure~\ref{fig:CaII_vs_NaI} indicates that at the LVA velocities the Ca\,\textsc{ii}\,K emission is very bright, with $L/C \approx 30$. Under the assumption that the LVA gas crosses the stellar continuum but misses the line emitter, we set $\rm{CF}_c\simeq 1$ and solve Eq.~\eqref{eq:CF_line_max} for $\rm{CF}_\ell$. The result is
        \begin{multline}
            \rm{CF}_{\ell} < \frac{\varepsilon(1+L/C)}{(L/C)\,(1-e^{-\tau})}\;-\;\frac{1}{L/C} \\
            \stackrel{L/C=30}{=}
        \frac{0.065\times 31}{30\,[1-e^{-\tau}]}\;-\;\frac{1}{30}.
        \end{multline}
        For representative Ca\,\textsc{ii} opacities at the LVA velocity of $\tau = 0.5$, $1$, and $2$, we find $\rm{CF}_{\ell} \lesssim 0.14$, $0.07$, and $0.05$.        
        Thus, with $L/C \approx 30$ the LVA remains undetected in  \ion{Ca}{ii} as long as it is projected against the line emitter with $\rm{CF}_{\ell} \sim 10\%$, even though it fully covers the continuum (\(\mathrm{CF}_c\approx 1\)).
        
        For the MVA, Fig.~\ref{fig:CaII_vs_NaI} shows that the absorption is deep relative to the line but the profile never falls below the continuum. This suggests that the absorbing gas fully covers the line-emitting region ($\rm{CF}_\ell \approx 1$) but is weakly projected against the continuum source ($\rm{CF}_c \ll 1$), so that $F(v)/C \approx 1 + (L(v)/C)e^{-\tau}$ and the absorbed flux remains above the continuum (i.e., $F(v)/C > 1$).
        In summary, this analysis suggests that the visibility contrast between LVA and MVA in \ion{Ca}{ii} is governed primarily by geometry and not by large differences in ionization.
    \end{appendix}

\end{document}